\DocumentMetadata{}
\documentclass[sigconf]{acmart}

\usepackage{graphicx}
\usepackage{textcomp}
\usepackage{xcolor}
\usepackage{xspace}

\usepackage{pgfplots}
\pgfplotsset{compat=1.10} 
\usepackage{float}
\usepackage{booktabs}
\RequirePackage{xstring}
\RequirePackage{xparse}
\RequirePackage{acro}
\usepackage{url}
\usepackage{pgfplots}
\usepgfplotslibrary{groupplots}
\usepackage{caption}
\usepackage{subcaption}
\usepackage[capitalise]{cleveref}
\crefname{section}{Sect.}{Sect.}
\crefname{figure}{Figure}{Fig.}
\crefname{table}{Tab.}{Tab.}
\crefname{equation}{Eq.}{Eq.}
\usepackage{tikz}
\usepackage{algorithm}
\usepackage{listings}
\usepackage{xcolor}
\usepackage[noend]{algorithmic}
\newlength\fheight
\newlength\fwidth
\usepackage{todonotes}
\usepackage{circledsteps}
\makeatletter

\def\BibTeX{{\rm B\kern-.05em{\sc i\kern-.025em b}\kern-.08em
    T\kern-.1667em\lower.7ex\hbox{E}\kern-.125emX}}
\usepackage[acronyms,nonumberlist,nopostdot,nomain,nogroupskip,acronymlists={hidden}]{glossaries}
\newglossary[algh]{hidden}{acrh}{acnh}{Hidden Acronyms}

\usepackage{soul}

\setcopyright{acmlicensed}
\copyrightyear{2026}
\acmYear{2026}
\acmDOI{XXXXXXX.XXXXXXX}
\acmConference[MobiHoc '26]{Make sure to enter the correct
  conference title from your rights confirmation email}{Nov 23–26, 2026}{Tokyo, Japan}
\acmISBN{978-1-4503-XXXX-X/2018/06}

\newacronym{3gpp}{3GPP}{3rd Generation Partnership Project}
\newacronym{4g}{4G}{4th generation}
\newacronym{5g}{5G}{5th generation}
\newacronym{6g}{6G}{6th generation}
\newacronym{5gc}{5GC}{5G Core}
\newacronym{adc}{ADC}{Analog to Digital Converter}
\newacronym{aerpaw}{AERPAW}{Aerial Experimentation and Research Platform for Advanced Wireless}
\newacronym{ai}{AI}{Artificial Intelligence}
\newacronym{aimd}{AIMD}{Additive Increase Multiplicative Decrease}
\newacronym{am}{AM}{Acknowledged Mode}
\newacronym{amc}{AMC}{Adaptive Modulation and Coding}
\newacronym{amf}{AMF}{Access and Mobility Management Function}
\newacronym{aops}{AOPS}{Adaptive Order Prediction Scheduling}
\newacronym{api}{API}{Application Programming Interface}
\newacronym{apn}{APN}{Access Point Name}
\newacronym{ap}{AP}{Application Protocol}
\newacronym{aqm}{AQM}{Active Queue Management}
\newacronym{ausf}{AUSF}{Authentication Server Function}
\newacronym{avc}{AVC}{Advanced Video Coding}
\newacronym{awgn}{AGWN}{Additive White Gaussian Noise}
\newacronym{balia}{BALIA}{Balanced Link Adaptation Algorithm}
\newacronym{bbu}{BBU}{Base Band Unit}
\newacronym{bdp}{BDP}{Bandwidth-Delay Product}
\newacronym{ber}{BER}{Bit Error Rate}
\newacronym{bf}{BF}{Beamforming}
\newacronym{bler}{BLER}{Block Error Rate}
\newacronym{brr}{BRR}{Bayesian Ridge Regressor}
\newacronym{bs}{BS}{Base Station}
\newacronym{bsr}{BSR}{Buffer Status Report}
\newacronym{bss}{BSS}{Business Support System}
\newacronym{ca}{CA}{Carrier Aggregation}
\newacronym{caas}{CaaS}{Connectivity-as-a-Service}
\newacronym{cb}{CB}{Code Block}
\newacronym{cc}{CC}{Congestion Control}
\newacronym{ccid}{CCID}{Congestion Control ID}
\newacronym{cco}{CC}{Carrier Component}
\newacronym{cd}{CD}{Continuous Delivery}
\newacronym{cdd}{CDD}{Cyclic Delay Diversity}
\newacronym{cdf}{CDF}{Cumulative Distribution Function}
\newacronym{cdn}{CDN}{Content Distribution Network}
\newacronym{cli}{CLI}{Command-line Interface}
\newacronym{cn}{CN}{Core Network}
\newacronym{codel}{CoDel}{Controlled Delay Management}
\newacronym{comac}{COMAC}{Converged Multi-Access and Core}
\newacronym{cord}{CORD}{Central Office Re-architected as a Datacenter}
\newacronym{cornet}{CORNET}{COgnitive Radio NETwork}
\newacronym{cosmos}{COSMOS}{Cloud Enhanced Open Software Defined Mobile Wireless Testbed for City-Scale Deployment}
\newacronym{cots}{COTS}{Commercial Off-the-Shelf}
\newacronym{cp}{CP}{Control Plane}
\newacronym{cpt}{CPT}{Conditional Probability Table}
\newacronym{cyp}{CP}{Cyclic Prefix}
\newacronym{up}{UP}{User Plane}
\newacronym{cpu}{CPU}{Central Processing Unit}
\newacronym{cqi}{CQI}{Channel Quality Information}
\newacronym{cr}{CR}{Cognitive Radio}
\newacronym{cran}{CRAN}{Cloud \gls{ran}}
\newacronym{crs}{CRS}{Cell Reference Signal}
\newacronym{csi}{CSI}{Channel State Information}
\newacronym{csirs}{CSI-RS}{Channel State Information - Reference Signal}
\newacronym{cu}{CU}{Central Unit}
\newacronym{d2tcp}{D$^2$TCP}{Deadline-aware Data center TCP}
\newacronym{d3}{D$^3$}{Deadline-Driven Delivery}
\newacronym{dac}{DAC}{Digital to Analog Converter}
\newacronym{dag}{DAG}{Directed Acyclic Graph}
\newacronym{das}{DAS}{Distributed Antenna System}
\newacronym{dash}{DASH}{Dynamic Adaptive Streaming over HTTP}
\newacronym{dc}{DC}{Dual Connectivity}
\newacronym{dccp}{DCCP}{Datagram Congestion Control Protocol}
\newacronym{dce}{DCE}{Direct Code Execution}
\newacronym{dci}{DCI}{Downlink Control Information}
\newacronym{dctcp}{DCTCP}{Data Center TCP}
\newacronym{dl}{DL}{Downlink}
\newacronym{dmr}{DMR}{Deadline Miss Ratio}
\newacronym{dmrs}{DMRS}{DeModulation Reference Signal}
\newacronym{drlcc}{DRL-CC}{Deep Reinforcement Learning Congestion Control}
\newacronym{drs}{DRS}{Discovery Reference Signal}
\newacronym{du}{DU}{Distributed Unit}
\newacronym{e2e}{E2E}{end-to-end}
\newacronym{earfcn}{EARFCN}{E-UTRA Absolute Radio Frequency Channel Number}
\newacronym{ecaas}{ECaaS}{Edge-Cloud-as-a-Service}
\newacronym{ecn}{ECN}{Explicit Congestion Notification}
\newacronym{edf}{EDF}{Earliest Deadline First}
\newacronym{embb}{eMBB}{Enhanced Mobile Broadband}
\newacronym{empower}{EMPOWER}{EMpowering transatlantic PlatfOrms for advanced WirEless Research}
\newacronym{enb}{eNB}{evolved Node Base}
\newacronym{endc}{EN-DC}{E-UTRAN-\gls{nr} \gls{dc}}
\newacronym{epc}{EPC}{Evolved Packet Core}
\newacronym{eps}{EPS}{Evolved Packet System}
\newacronym{es}{ES}{Edge Server}
\newacronym{etsi}{ETSI}{European Telecommunications Standards Institute}
\newacronym[firstplural=Estimated Times of Arrival (ETAs)]{eta}{ETA}{Estimated Time of Arrival}
\newacronym{eutran}{E-UTRAN}{Evolved Universal Terrestrial Access Network}
\newacronym{faas}{FaaS}{Function-as-a-Service}
\newacronym{fapi}{FAPI}{Functional Application Platform Interface}
\newacronym{fdd}{FDD}{Frequency Division Duplexing}
\newacronym{fdm}{FDM}{Frequency Division Multiplexing}
\newacronym{fdma}{FDMA}{Frequency Division Multiple Access}
\newacronym{fed4fire}{FED4FIRE+}{Federation 4 Future Internet Research and Experimentation Plus}
\newacronym{fir}{FIR}{Finite Impulse Response}
\newacronym{fit}{FIT}{Future \acrlong{iot}}
\newacronym{fpga}{FPGA}{Field Programmable Gate Array}
\newacronym{fr2}{FR2}{Frequency Range 2}
\newacronym{fs}{FS}{Fast Switching}
\newacronym{fscc}{FSCC}{Flow Sharing Congestion Control}
\newacronym{ftp}{FTP}{File Transfer Protocol}
\newacronym{fw}{FW}{Flow Window}
\newacronym{ge}{GE}{Gaussian Elimination}
\newacronym{gnb}{gNB}{Next Generation Node Base}
\newacronym{llm}{LLM}{Large Language Model}
\newacronym{gop}{GOP}{Group of Pictures}
\newacronym{gpr}{GPR}{Gaussian Process Regressor}
\newacronym{gpu}{GPU}{Graphics Processing Unit}
\newacronym{gtp}{GTP}{GPRS Tunneling Protocol}
\newacronym{gtpc}{GTP-C}{GPRS Tunnelling Protocol Control Plane}
\newacronym{gtpu}{GTP-U}{GPRS Tunnelling Protocol User Plane}
\newacronym{gtpv2c}{GTPv2-C}{\gls{gtp} v2 - Control}
\newacronym{fwa}{FWA}{Fixed Wireless Access}
\newacronym{bic}{BIC}{Bayesian Information Criterion}
\newacronym{gw}{GW}{Gateway}
\newacronym{harq}{HARQ}{Hybrid Automatic Repeat reQuest}
\newacronym{hetnet}{HetNet}{Heterogeneous Network}
\newacronym{hh}{HH}{Hard Handover}
\newacronym{hol}{HOL}{Head-of-Line}
\newacronym{hqf}{HQF}{Highest-quality-first}
\newacronym{hss}{HSS}{Home Subscription Server}
\newacronym{http}{HTTP}{HyperText Transfer Protocol}
\newacronym{ia}{IA}{Initial Access}
\newacronym{iab}{IAB}{Integrated Access and Backhaul}
\newacronym{ic}{IC}{Incident Command}
\newacronym{ietf}{IETF}{Internet Engineering Task Force}
\newacronym{imsi}{IMSI}{International Mobile Subscriber Identity}
\newacronym{imt}{IMT}{International Mobile Telecommunication}
\newacronym{iot}{IoT}{Internet of Things}
\newacronym{ip}{IP}{Internet Protocol}
\newacronym{itu}{ITU}{International Telecommunication Union}
\newacronym{kpi}{KPI}{Key Performance Indicator}
\newacronym{kpm}{KPM}{Key Performance Measurement}
\newacronym{kvm}{KVM}{Kernel-based Virtual Machine}
\newacronym{los}{LoS}{Line of Sight}
\newacronym{lsm}{LSM}{Link-to-System Mapping}
\newacronym{lstm}{LSTM}{Long Short Term Memory}
\newacronym{lte}{LTE}{Long Term Evolution}
\newacronym{lxc}{LXC}{Linux Container}
\newacronym{m2m}{M2M}{Machine to Machine}
\newacronym{mac}{MAC}{Medium Access Control}
\newacronym{manet}{MANET}{Mobile Ad Hoc Network}
\newacronym{mano}{MANO}{Management and Orchestration}
\newacronym{mc}{MC}{Multi-Connectivity}
\newacronym{mcc}{MCC}{Mobile Cloud Computing}
\newacronym{mchem}{MCHEM}{Massive Channel Emulator}
\newacronym{mcs}{MCS}{Modulation and Coding Scheme}
\newacronym{mec2}{MEC}{Multi-access Edge Computing}
\newacronym{mec}{MEC}{Mobile Edge Computing}
\newacronym{mfc}{MFC}{Mobile Fog Computing}
\newacronym{mgen}{MGEN}{Multi-Generator}
\newacronym{mi}{MI}{Mutual Information}
\newacronym{mib}{MIB}{Master Information Block}
\newacronym{miesm}{MIESM}{Mutual Information Based Effective SINR}
\newacronym{mimo}{MIMO}{Multiple Input, Multiple Output}
\newacronym{ml}{ML}{Machine Learning}
\newacronym{mlr}{MLR}{Maximum-local-rate}
\newacronym[plural=\gls{mme}s,firstplural=Mobility Management Entities (MMEs)]{mme}{MME}{Mobility Management Entity}
\newacronym{mmtc}{mMTC}{Massive Machine-Type Communications}
\newacronym{mmwave}{mmWave}{millimeter wave}
\newacronym{mpdccp}{MP-DCCP}{Multipath Datagram Congestion Control Protocol}
\newacronym{mptcp}{MPTCP}{Multipath TCP}
\newacronym{mr}{MR}{Maximum Rate}
\newacronym{bdeu}{BDeU}{Bayesian Dirichlet equivalent uniform}
\newacronym{mrdc}{MR-DC}{Multi \gls{rat} \gls{dc}}
\newacronym{mse}{MSE}{Mean Square Error}
\newacronym{mss}{MSS}{Maximum Segment Size}
\newacronym{mt}{MT}{Mobile Termination}
\newacronym{mtd}{MTD}{Machine-Type Device}
\newacronym{mtu}{MTU}{Maximum Transmission Unit}
\newacronym{mumimo}{MU-MIMO}{Multi-user \gls{mimo}}
\newacronym{mvno}{MVNO}{Mobile Virtual Network Operator}
\newacronym{nalu}{NALU}{Network Abstraction Layer Unit}
\newacronym{nas}{NAS}{Network Attached Storage}
\newacronym{nat}{NAT}{Network Address Translation}
\newacronym{nbiot}{NB-IoT}{Narrow Band IoT}
\newacronym{nfv}{NFV}{Network Function Virtualization}
\newacronym{nfvi}{NFVI}{Network Function Virtualization Infrastructure}
\newacronym{ni}{NI}{Network Interfaces}
\newacronym{nic}{NIC}{Network Interface Card}
\newacronym{now}{NOW}{Non Overlapping Window}
\newacronym{cpd}{CPD}{Conditional Probability Distribution}
\newacronym{nsm}{NSM}{Network Service Mesh}
\newacronym{nr}{NR}{New Radio}
\newacronym{nrf}{NRF}{Network Repository Function}
\newacronym{nsa}{NSA}{Non Stand Alone}
\newacronym{nse}{NSE}{Network Slicing Engine}
\newacronym{nssf}{NSSF}{Network Slice Selection Function}
\newacronym{o2i}{O2I}{Outdoor to Indoor}
\newacronym{oai}{OAI}{OpenAirInterface}
\newacronym{oaicn}{OAI-CN}{\gls{oai} \acrlong{cn}}
\newacronym{oairan}{OAI-RAN}{\acrlong{oai} \acrlong{ran}}
\newacronym{oam}{OAM}{Operations, Administration and Maintenance}
\newacronym{ofdm}{OFDM}{Orthogonal Frequency Division Multiplexing}
\newacronym{olia}{OLIA}{Opportunistic Linked Increase Algorithm}
\newacronym{omec}{OMEC}{Open Mobile Evolved Core}
\newacronym{onap}{ONAP}{Open Network Automation Platform}
\newacronym{onf}{ONF}{Open Networking Foundation}
\newacronym{onos}{ONOS}{Open Networking Operating System}
\newacronym{oom}{OOM}{\gls{onap} Operations Manager}
\newacronym{opnfv}{OPNFV}{Open Platform for \gls{nfv}}
\newacronym{oran}{O-RAN}{Open Radio Access Network}
\newacronym{orbit}{ORBIT}{Open-Access Research Testbed for Next-Generation Wireless Networks}
\newacronym{os}{OS}{Operating System}
\newacronym{oss}{OSS}{Operations Support System}
\newacronym{pa}{PA}{Position-aware}
\newacronym{pase}{PASE}{Prioritization, Arbitration, and Self-adjusting Endpoints}
\newacronym{pawr}{PAWR}{Platforms for Advanced Wireless Research}
\newacronym{pbch}{PBCH}{Physical Broadcast Channel}
\newacronym{pcef}{PCEF}{Policy and Charging Enforcement Function}
\newacronym{pcfich}{PCFICH}{Physical Control Format Indicator Channel}
\newacronym{pcrf}{PCRF}{Policy and Charging Rules Function}
\newacronym{pdcch}{PDCCH}{Physical Downlink Control Channel}
\newacronym{pdcp}{PDCP}{Packet Data Convergence Protocol}
\newacronym{pdsch}{PDSCH}{Physical Downlink Shared Channel}
\newacronym{pdu}{PDU}{Packet Data Unit}
\newacronym{pf}{PF}{Proportional Fair}
\newacronym{pgw}{PGW}{Packet Gateway}
\newacronym{phich}{PHICH}{Physical Hybrid ARQ Indicator Channel}
\newacronym{phy}{PHY}{Physical}
\newacronym{pmch}{PMCH}{Physical Multicast Channel}
\newacronym{pmi}{PMI}{Precoding Matrix Indicators}
\newacronym{powder}{POWDER}{Platform for Open Wireless Data-driven Experimental Research}
\newacronym{ppo}{PPO}{Proximal Policy Optimization}
\newacronym{ppp}{PPP}{Poisson Point Process}
\newacronym{prach}{PRACH}{Physical Random Access Channel}
\newacronym{prb}{PRB}{Physical Resource Block}
\newacronym{psnr}{PSNR}{Peak Signal to Noise Ratio}
\newacronym{pss}{PSS}{Primary Synchronization Signal}
\newacronym{pucch}{PUCCH}{Physical Uplink Control Channel}
\newacronym{pusch}{PUSCH}{Physical Uplink Shared Channel}
\newacronym{qam}{QAM}{Quadrature Amplitude Modulation}
\newacronym{qci}{QCI}{\gls{qos} Class Identifier}
\newacronym{qoe}{QoE}{Quality of Experience}
\newacronym{qos}{QoS}{Quality of Service}
\newacronym{quic}{QUIC}{Quick UDP Internet Connections}
\newacronym{rach}{RACH}{Random Access Channel}
\newacronym{ran}{RAN}{Radio Access Network}
\newacronym[firstplural=Radio Access Technologies (RATs)]{rat}{RAT}{Radio Access Technology}
\newacronym{rbg}{RBG}{Resource Block Group}
\newacronym{rcn}{RCN}{Research Coordination Network}
\newacronym{rc}{RC}{RAN Control}
\newacronym{rec}{REC}{Radio Edge Cloud}
\newacronym{red}{RED}{Random Early Detection}
\newacronym{renew}{RENEW}{Reconfigurable Eco-system for Next-generation End-to-end Wireless}
\newacronym{rf}{RF}{Radio Frequency}
\newacronym{rfc}{RFC}{Request for Comments}
\newacronym{rfr}{RFR}{Random Forest Regressor}
\newacronym{ric}{RIC}{RAN Intelligent Controller}
\newacronym{rlc}{RLC}{Radio Link Control}
\newacronym{rlf}{RLF}{Radio Link Failure}
\newacronym{rlnc}{RLNC}{Random Linear Network Coding}
\newacronym{rmr}{RMR}{RIC Message Router}
\newacronym{rmse}{RMSE}{Root Mean Squared Error}
\newacronym{rnis}{RNIS}{Radio Network Information Service}
\newacronym{rr}{RR}{Round Robin}
\newacronym{rrc}{RRC}{Radio Resource Control}
\newacronym{rrm}{RRM}{Radio Resource Management}
\newacronym{rru}{RRU}{Remote Radio Unit}
\newacronym{rs}{RS}{Remote Server}
\newacronym{rsrp}{RSRP}{Reference Signal Received Power}
\newacronym{rsrq}{RSRQ}{Reference Signal Received Quality}
\newacronym{rss}{RSS}{Received Signal Strength}
\newacronym{rssi}{RSSI}{Received Signal Strength Indicator}
\newacronym{rtt}{RTT}{Round Trip Time}
\newacronym{ru}{RU}{Radio Unit}
\newacronym{rw}{RW}{Receive Window}
\newacronym{rx}{RX}{Receiver}
\newacronym{s1ap}{S1AP}{S1 Application Protocol}
\newacronym{sa}{SA}{standalone}
\newacronym{sack}{SACK}{Selective Acknowledgment}
\newacronym{sap}{SAP}{Service Access Point}
\newacronym{sc2}{SC2}{Spectrum Collaboration Challenge}
\newacronym{scef}{SCEF}{Service Capability Exposure Function}
\newacronym{sch}{SCH}{Secondary Cell Handover}
\newacronym{scoot}{SCOOT}{Split Cycle Offset Optimization Technique}
\newacronym{sctp}{SCTP}{Stream Control Transmission Protocol}
\newacronym{sdap}{SDAP}{Service Data Adaptation Protocol}
\newacronym{sdk}{SDK}{Software Development Kit}
\newacronym{sdm}{SDM}{Space Division Multiplexing}
\newacronym{sdma}{SDMA}{Spatial Division Multiple Access}
\newacronym{sdn}{SDN}{Software-defined Networking}
\newacronym{sdr}{SDR}{Software-defined Radio}
\newacronym{seba}{SEBA}{SDN-Enabled Broadband Access}
\newacronym{sgsn}{SGSN}{Serving GPRS Support Node}
\newacronym{sgw}{SGW}{Service Gateway}
\newacronym{si}{SI}{Study Item}
\newacronym{sib}{SIB}{Secondary Information Block}
\newacronym{sinr}{SINR}{Signal to Interference plus Noise Ratio}
\newacronym{sip}{SIP}{Session Initiation Protocol}
\newacronym{siso}{SISO}{Single Input, Single Output}
\newacronym{sla}{SLA}{Service Level Agreement}
\newacronym{sm}{SM}{Service Model}
\newacronym{smo}{SMO}{Service Management and Orchestration}
\newacronym{smsgmsc}{SMS-GMSC}{\gls{sms}-Gateway}
\newacronym{snr}{SNR}{Signal-to-Noise-Ratio}
\newacronym{son}{SON}{Self-Organizing Network}
\newacronym{sptcp}{SPTCP}{Single Path TCP}
\newacronym{srb}{SRB}{Service Radio Bearer}
\newacronym{srn}{SRN}{Standard Radio Node}
\newacronym{srs}{SRS}{Sounding Reference Signal}
\newacronym{ss}{SS}{Synchronization Signal}
\newacronym{sss}{SSS}{Secondary Synchronization Signal}
\newacronym{st}{ST}{Spanning Tree}
\newacronym{svc}{SVC}{Scalable Video Coding}
\newacronym{tb}{TB}{Transport Block}
\newacronym{tcp}{TCP}{Transmission Control Protocol}
\newacronym{tdd}{TDD}{Time Division Duplexing}
\newacronym{tdm}{TDM}{Time Division Multiplexing}
\newacronym{tdma}{TDMA}{Time Division Multiple Access}
\newacronym{tfl}{TfL}{Transport for London}
\newacronym{tfrc}{TFRC}{TCP-Friendly Rate Control}
\newacronym{tft}{TFT}{Traffic Flow Template}
\newacronym{tgen}{TGEN}{Traffic Generator}
\newacronym{tip}{TIP}{Telecom Infra Project}
\newacronym{tm}{TM}{Transparent Mode}
\newacronym{to}{TO}{Telco Operator}
\newacronym{tr}{TR}{Technical Report}
\newacronym{trp}{TRP}{Transmitter Receiver Pair}
\newacronym{ts}{TS}{Technical Specification}
\newacronym{tti}{TTI}{Transmission Time Interval}
\newacronym{ttt}{TTT}{Time-to-Trigger}
\newacronym{tx}{TX}{Transmitter}
\newacronym{uas}{UAS}{Unmanned Aerial System}
\newacronym{uav}{UAV}{Unmanned Aerial Vehicle}
\newacronym{udm}{UDM}{Unified Data Management}
\newacronym{udp}{UDP}{User Datagram Protocol}
\newacronym{udr}{UDR}{Unified Data Repository}
\newacronym{ue}{UE}{User Equipment}
\newacronym{uhd}{UHD}{\gls{usrp} Hardware Driver}
\newacronym{ul}{UL}{Uplink}
\newacronym{um}{UM}{Unacknowledged Mode}
\newacronym{uml}{UML}{Unified Modeling Language}
\newacronym{upa}{UPA}{Uniform Planar Array}
\newacronym{upf}{UPF}{User Plane Function}
\newacronym{urllc}{URLLC}{Ultra Reliable and Low Latency Communications}
\newacronym{usa}{U.S.}{United States}
\newacronym{usim}{USIM}{Universal Subscriber Identity Module}
\newacronym{usrp}{USRP}{Universal Software Radio Peripheral}
\newacronym{utc}{UTC}{Urban Traffic Control}
\newacronym{vim}{VIM}{Virtualization Infrastructure Manager}
\newacronym{vm}{VM}{Virtual Machine}
\newacronym{vnf}{VNF}{Virtual Network Function}
\newacronym{volte}{VoLTE}{Voice over \gls{lte}}
\newacronym{voltha}{VOLTHA}{Virtual OLT HArdware Abstraction}
\newacronym{vr}{VR}{Virtual Reality}
\newacronym{vran}{vRAN}{Virtualized \gls{ran}}
\newacronym{vss}{VSS}{Video Streaming Server}
\newacronym{wbf}{WBF}{Wired Bias Function}
\newacronym{wf}{WF}{Waterfilling}
\newacronym{wg}{WG}{Working Group}
\newacronym{wlan}{WLAN}{Wireless Local Area Network}
\newacronym{osm}{OSM}{Open Source \gls{nfv} Management and Orchestration}
\newacronym{pnf}{PNF}{Physical Network Function}
\newacronym{drl}{DRL}{Deep Reinforcement Learning}
\newacronym{mtc}{MTC}{Machine-type Communications}
\newacronym{osc}{OSC}{O-RAN Software Community}
\newacronym{mns}{MnS}{Management Services}
\newacronym{ves}{VES}{\gls{vnf} Event Stream}
\newacronym{ei}{EI}{Enrichment Information}
\newacronym{fh}{FH}{Fronthaul}
\newacronym{fft}{FFT}{Fast Fourier Transform}
\newacronym{laa}{LAA}{Licensed-Assisted Access}
\newacronym{plfs}{PLFS}{Physical Layer Frequency Signals}
\newacronym{ptp}{PTP}{Precision Time Protocol}
\newacronym{lidar}{LiDAR}{Light Detection And Ranging}
\newacronym{dem}{DEM}{Digital Elevation Model}
\newacronym{dtm}{DEM}{Digital Terrain Model}
\newacronym{dsm}{DEM}{Digital Surface Models}
\newacronym{ota}{OTA}{Over-The-Air}
\newacronym{ns}{NS}{Network Slicing}
\newacronym{ne}{NE}{Nash Equilibrium}
\newacronym{hf}{HF}{High Frequency}
\newacronym{noma}{NOMA}{Non-Orthogonal Multiple Access}
\newacronym{sre}{SRE}{Smart Radio Environment}
\newacronym{ris}{RIS}{Reconfigurable Intelligent Surface}
\newacronym{inp}{InP}{Infrastructure Provider}
\newacronym{smf}{SMF}{Slicing Magangement Framework}
\newacronym{nsn}{NSN}{Network Slicing Negotiation}
\newacronym{sms}{SMS}{Slicing MAC Scheduler}
\newacronym{brd}{BRD}{Best Response Dynamics}
\newacronym{dssbr}{DSSBR}{Double Step Smoothed Best Response}
\newacronym{poa}{PoA}{Price of Anarchy}
\newacronym{pos}{PoS}{Price of Stability}
\newacronym{milp}{MILP}{Mixed Integer-Linear Program}
\newacronym{pod}{PoD}{Price of DSSBR}
\newacronym{roc}{ROC}{Radio Overload Control}
\newacronym{ciot}{cIoT}{critical Internet of Things}
\newacronym{embbpr}{eMBB Pr.}{enhanced Mobile BroadBand Premium}
\newacronym{embbbs}{eMBB Bs.}{enhanced Mobile BroadBand Basic}
\newacronym{en}{EN}{Edge Node}
\newacronym{ec}{EC}{Edge Computing}
\newacronym{sp}{SP}{Service Provider}
\newacronym{me}{ME}{Market Equilibrium}
\newacronym{so}{SO}{Social Optimum}
\newacronym{wso}{WSO}{Weighted Social Optimum}
\newacronym{wsn}{WSN}{Wireless Sensor Network}
\newacronym{ps}{PS}{Proportional Sharing}
\newacronym{eg}{EG}{Eisenberg-Gale program}
\newacronym{pe}{PE}{Pareto Efficiency}
\newacronym{nsw}{NSW}{Nash Social Welfare}
\newacronym{ef}{EF}{Envy-Freeness}
\newacronym{sub6}{sub6GHz}{Below 6GHz}
\newacronym{ncr}{NCR}{Network-Controlled Repeater}
\newacronym{nlos}{NLoS}{Non-LoS}
\newacronym{src}{SRC}{Smart Radio Connection}
\newacronym{srd}{SRD}{Smart Radio Device}
\newacronym{cs}{CS}{Candidate Site}
\newacronym{tp}{TP}{Test Point}
\newacronym{fov}{FoV}{Field of View}
\newacronym{nrric}{Near-RT RIC}{Near Real-time RAN Intelligent Controller}
\newacronym{e2ap}{E2AP}{E2 Application Protocol}
\newacronym{e2sm}{E2SM}{E2 Service Model}
\newacronym{nrtric}{Non-RT RIC}{Non-Real-Time Ran Intelligent Controller}
\newacronym{itti}{ITTI}{Inter-task Interface}
\newacronym{bap}{BAP}{Backhaul Adaptation Protocol}
\newacronym{iabest}{IABEST}{Integrated Access and Backhaul Experimental large-Scale Tetbed}
\newacronym{teid}{TEID}{Tunnel Endpoint Identifier}
\newacronym{dlsch}{DL-SCH}{Downlink Shared Channel }
\newacronym{ulsch}{UL-SCH}{Uplink Shared Channel }
\newacronym{opex}{OpEx}{Operational Expenditure}
\newacronym{capex}{CapEx}{Capital Expenditure}
\newacronym{mno}{MNO}{Mobile Network Operator}
\newacronym{fr}{FR}{Frequency Range}
\newacronym{bn}{BN}{Bayesian Network}
\newcommand{\framework}{BLINC\xspace}

\title{\framework: Context-Specific Causal Learning for Automated RAN Configuration}

\begin{abstract}
\gls{ran} configuration has traditionally required significant manual effort due to 
indirect causal dependencies between observable \glspl{kpi}, and context-dependent characteristics, where the optimal configurations vary with network conditions. Although recent data-driven approaches improve parameter tuning, they remain limited in distinguishing causal direction from statistical correlation and in generalizing across diverse operating contexts.

To address these challenges, we propose \framework (Bayesian \gls{llm}-Driven Intelligent Network Configuration),
an \gls{llm}-assisted Bayesian Network framework that integrates telecommunications domain knowledge into causal structure learning. 
Trained and validated on a private 5G deployment, our method achieves throughput improvement of 63.5\% with 19.7\% reduction on block error rate over data-only baselines through joint optimization of power control and link adaptation parameters. The framework provides interpretable causal structure, while also quantifying prediction uncertainty. We also demonstrate  the ability of the Bayesian Network framework to adapt to different deployment scenarios and propose an incremental \gls{cpd} update mechanism with learning rate for continuous model adaptation as network conditions evolve.
\end{abstract}


\begin{document}

\author{Reshma Prasad, Michele Polese, Tommaso Melodia}
\email{{re.prasad,  m.polese, melodia}@northeastern.edu.}
\affiliation{%
  \institution{Institute for Intelligent Networked Systems, Northeastern University}
  \city{Boston}
  \country{USA}
}

\captionsetup[subfigure]{font=scriptsize}

\maketitle
\begin{tikzpicture}[remember picture, overlay]
  \node[anchor=north, font=\small,  text width=\paperwidth-1cm, align=center] 
    at ([yshift=-1.0cm]current page.north) 
    {This work has been submitted to the ACM for possible publication. 
     Copyright may be transferred without notice, after which this version 
     may no longer be accessible.};
\end{tikzpicture}
\glsresetall
\glsunset{oran}

\section{Introduction}

The evolution toward \gls{6g} networks is driven by emerging applications and deployment scenarios with diverse \gls{qos} requirements~\cite{tataria20216g}. 
These must be met across highly heterogeneous environments---from indoor private networks to outdoor macro sites in urban and rural areas---all served by the same \gls{ran} protocol stack.
Effective \gls{ran} management is therefore essential to deliver such diverse applications from the same infrastructure. 
In particular, optimizing \gls{ran} configuration parameters is critical, as proper tuning directly impacts the performance experienced by end users.

The relationships between \gls{ran} configuration parameters and \glspl{kpi}, however, are inherently complex and indirect. Changes in network configurations propagate through multiple intermediate variables before affecting observable \glspl{kpi}. Therefore, there are multiple layers of dependencies and feedback mechanisms between configurations and \glspl{kpi}.
Evaluating and optimizing the effects on performance of configuration changes in real deployments demands extensive domain expertise and practical experience~\cite{ge2025iridescence}. For example, the \gls{oai} \gls{gnb} stack has approximately 70 tunable parameters that include frequency and bandwidth settings, power control, \gls{rach} configuration, \gls{tdd} frame structure, \gls{mac}/\gls{rlc} layer settings,  and system level configurations.
Manually deriving causal relationships between dozens of configuration parameters and multiple \glspl{kpi} presents significant scalability challenges. In this work, we focus on uplink power control parameters as a controlled, rigorous validation of the framework. The framework is designed to extend to broader parameter sets, leveraging the incremental update mechanism to progressively refine causal structures.
\begin{figure}[tbp]
\begin{subfigure}[t]{0.48\columnwidth}
    \centering
            \setlength\fwidth{\linewidth}
            \setlength\fheight{0.8\linewidth}
            \begin{tikzpicture}

\definecolor{darkgray176}{RGB}{176,176,176}
\definecolor{dodgerblue52152219}{RGB}{52,152,219}
\definecolor{lightgray204}{RGB}{204,204,204}
\definecolor{mediumseagreen46204113}{RGB}{46,204,113}
\definecolor{orange24315618}{RGB}{243,156,18}
\definecolor{tomato2317660}{RGB}{231,76,60}


\begin{axis}[
name=main,
width=\fwidth,
height=\fheight,
legend cell align={left},
legend columns=3,
legend style={
  fill opacity=0.8,
  draw opacity=1,
  text opacity=1,
  at={(0.5,1.03)},
  anchor=south,
  draw=gray,
  font=\scriptsize
},
tick align=inside,
tick pos=left,
axis y line*=left,
x grid style={darkgray176},
xmin=-0.22, xmax=0.72,
xtick style={color=black},
xtick={0,0.5},
xticklabels={FWA UE,Mobility UE},
y grid style={darkgray176},
ylabel={Throughput (Mbps)},
ylabel shift={-5pt},
ymajorgrids,
ymin=0, ymax=70,
ytick style={color=black},
font=\scriptsize
]

\draw[draw=black,fill=mediumseagreen46204113,opacity=0.8,thick] (axis cs:-0.17,0) rectangle (axis cs:-0.07,51.559049086758);
\addlegendimage{ybar,ybar legend,draw=black,fill=mediumseagreen46204113,opacity=0.8,thick}
\addlegendentry{Throughput}

\draw[draw=black,fill=mediumseagreen46204113,opacity=0.8,thick] (axis cs:0.33,0) rectangle (axis cs:0.43,8.68233644859813);

\draw[draw=black,fill=orange24315618,opacity=0.8,thick] (axis cs:-0.05,0) rectangle (axis cs:0.05,14.3315068493151);
\addlegendimage{ybar,ybar legend,draw=black,fill=orange24315618,opacity=0.8,thick}
\addlegendentry{CQI}

\draw[draw=black,fill=orange24315618,opacity=0.8,thick] (axis cs:0.45,0) rectangle (axis cs:0.55,6.89096573208723);

\addlegendimage{ybar,ybar legend,draw=black,fill=dodgerblue52152219,opacity=0.8,thick}
\addlegendentry{RSRP}

\path [draw=black, semithick]
(axis cs:-0.12,43.4451200376132)
--(axis cs:-0.12,59.6729781359028);

\path [draw=black, semithick]
(axis cs:0.38,-4.87611722748339)
--(axis cs:0.38,22.2407901246797);

\addplot [semithick, black, mark=-, mark size=5, mark options={solid}, only marks]
table {%
-0.12 43.4451200376132
0.38 -4.87611722748339
};
\addplot [semithick, black, mark=-, mark size=5, mark options={solid}, only marks]
table {%
-0.12 59.6729781359028
0.38 22.2407901246797
};

\path [draw=black, semithick]
(axis cs:0,12.9689632035487)
--(axis cs:0,15.6940504950814);

\path [draw=black, semithick]
(axis cs:0.5,3.37440500078355)
--(axis cs:0.5,10.4075264633909);

\addplot [semithick, black, mark=-, mark size=5, mark options={solid}, only marks]
table {%
0 12.9689632035487
0.5 3.37440500078355
};
\addplot [semithick, black, mark=-, mark size=5, mark options={solid}, only marks]
table {%
0 15.6940504950814
0.5 10.4075264633909
};

\draw (axis cs:-0.12,61.6729781359028) node[
  scale=0.6,
  anchor=base,
  text=black,
  rotate=0.0
]{\bfseries 51.6};
\draw (axis cs:0.38,24.2407901246797) node[
  scale=0.6,
  anchor=base,
  text=black,
  rotate=0.0
]{\bfseries 8.7};

\draw (axis cs:0,17.6940504950814) node[
  scale=0.6,
  anchor=base,
  text=black,
  rotate=0.0
]{\bfseries 14.3};
\draw (axis cs:0.5,12.4075264633909) node[
  scale=0.6,
  anchor=base,
  text=black,
  rotate=0.0
]{\bfseries 6.9};
\end{axis}

\begin{axis}[
width=\fwidth,
height=\fheight,
axis y line*=right,
axis x line=none,
xmin=-0.22, xmax=0.72,
xtick=\empty,
ylabel={RSRP (dBm)},
ylabel near ticks,
ymax=0, ymin=-120,
ytick style={draw=none},
font=\scriptsize,
ylabel shift={-5pt},
yticklabel style={xshift=-1pt},
]

\draw[draw=black,fill=dodgerblue52152219,opacity=0.8,thick] (axis cs:0.07,0) rectangle (axis cs:0.17,-68.4125570776256);
\draw[draw=black,fill=dodgerblue52152219,opacity=0.8,thick] (axis cs:0.57,0) rectangle (axis cs:0.67,-89.7102803738318);

\path [draw=black, semithick]
(axis cs:0.12,-62.7991097375109)
--(axis cs:0.12,-74.0260044177402);

\path [draw=black, semithick]
(axis cs:0.62,-86.313565014305)
--(axis cs:0.62,-93.1069957333585);

\addplot [semithick, black, mark=-, mark size=5, mark options={solid}, only marks]
table {%
0.12 -62.7991097375109
0.62 -86.313565014305
};
\addplot [semithick, black, mark=-, mark size=5, mark options={solid}, only marks]
table {%
0.12 -74.0260044177402
0.62 -93.1069957333585
};

\draw (axis cs:0.12,-76.0260044177402) node[
  scale=0.6,
  anchor=north,
  text=black,
  rotate=0.0
]{\bfseries -68.4};
\draw (axis cs:0.62,-95.1069957333585) node[
  scale=0.6,
  anchor=north,
  text=black,
  rotate=0.0
]{\bfseries -89.7};

\end{axis}

\end{tikzpicture}%
            \vspace{-.25cm}
            \caption{System performance with same configuration in different deployment scenarios.}
            \label{fig:baseintro}
\end{subfigure}
\begin{subfigure}[t]{0.48\columnwidth}
            \setlength\fwidth{\linewidth}%
            \setlength\fheight{0.8\linewidth}%
            \begin{tikzpicture}

\definecolor{darkgray176}{RGB}{176,176,176}
\definecolor{dodgerblue52152219}{RGB}{52,152,219}
\definecolor{lightgray204}{RGB}{204,204,204}
\definecolor{mediumseagreen46204113}{RGB}{46,204,113}
\definecolor{orange24315618}{RGB}{243,156,18}
\definecolor{tomato2317660}{RGB}{231,76,60}


\begin{axis}[
name=main,
width=\fwidth,
height=\fheight,
legend cell align={left},
legend columns=3,
legend style={
  fill opacity=0.8,
  draw opacity=1,
  text opacity=1,
  at={(0.5,1.03)},
  anchor=south,
  draw=gray,
  font=\scriptsize
},
tick align=inside,
tick pos=left,
axis y line*=left,
x grid style={darkgray176},
xmin=-0.22, xmax=0.72,
xtick style={color=black},
xtick={0,0.5},
xticklabels={Current config,Predicted config},
y grid style={darkgray176},
ylabel={Throughput (Mbps)},
ylabel shift={-5pt},
ymajorgrids,
ymin=0, ymax=40,
ytick style={color=black},
font=\scriptsize
]

\draw[draw=black,fill=mediumseagreen46204113,opacity=0.8,thick] (axis cs:-0.17,0) rectangle (axis cs:-0.07,4.18528169014085);
\addlegendimage{ybar,ybar legend,draw=black,fill=mediumseagreen46204113,opacity=0.8,thick}
\addlegendentry{Throughput}

\draw[draw=black,fill=mediumseagreen46204113,opacity=0.8,thick] (axis cs:0.33,0) rectangle (axis cs:0.43,24.2250694444444);

\draw[draw=black,fill=orange24315618,opacity=0.8,thick] (axis cs:-0.05,0) rectangle (axis cs:0.05,8.74647887323944);
\addlegendimage{ybar,ybar legend,draw=black,fill=orange24315618,opacity=0.8,thick}
\addlegendentry{CQI}

\draw[draw=black,fill=orange24315618,opacity=0.8,thick] (axis cs:0.45,0) rectangle (axis cs:0.55,7.27083333333333);

\addlegendimage{ybar,ybar legend,draw=black,fill=dodgerblue52152219,opacity=0.8,thick}
\addlegendentry{RSRP}

\path [draw=black, semithick]
(axis cs:-0.12,3.63101430752175)
--(axis cs:-0.12,4.73954907275994);

\path [draw=black, semithick]
(axis cs:0.38,20.1061039289029)
--(axis cs:0.38,28.344034959986);

\addplot [semithick, black, mark=-, mark size=5, mark options={solid}, only marks]
table {%
-0.12 3.63101430752175
0.38 20.1061039289029
};
\addplot [semithick, black, mark=-, mark size=5, mark options={solid}, only marks]
table {%
-0.12 4.73954907275994
0.38 28.344034959986
};

\path [draw=black, semithick]
(axis cs:0,8.22141055997625)
--(axis cs:0,9.27154718650262);

\path [draw=black, semithick]
(axis cs:0.5,5.5358830609761)
--(axis cs:0.5,9.00578360569057);

\addplot [semithick, black, mark=-, mark size=5, mark options={solid}, only marks]
table {%
0 8.22141055997625
0.5 5.5358830609761
};
\addplot [semithick, black, mark=-, mark size=5, mark options={solid}, only marks]
table {%
0 9.27154718650262
0.5 9.00578360569057
};

\draw (axis cs:-0.12,6.73954907275994) node[
  scale=0.6,
  text=black,
  rotate=0.0
]{\bfseries 4.2};
\draw (axis cs:0.38,30.344034959986) node[
  scale=0.6,
  anchor=base,
  text=black,
  rotate=0.0
]{\bfseries 24.2};

\draw (axis cs:0,11.2715471865026) node[
  scale=0.6,
  anchor=base,
  text=black,
  rotate=0.0
]{\bfseries 8.7};
\draw (axis cs:0.5,11.0057836056906) node[
  scale=0.6,
  anchor=base,
  text=black,
  rotate=0.0
]{\bfseries 7.3};

\draw[->, thick, black] (axis cs:0.15,4) -- (axis cs:0.31,4);
\node at (axis cs:0.23,7) {\scriptsize \textbf{\framework}};

\end{axis}

\begin{axis}[
width=\fwidth,
height=\fheight,
axis y line*=right,
axis x line=none,
xmin=-0.22, xmax=0.72,
xtick=\empty,
ylabel={RSRP (dBm)},
ylabel near ticks,
ymax=0, ymin=-120,
ytick style={draw=none},
font=\scriptsize,
ylabel shift={-5pt},
yticklabel style={xshift=-1pt},
]

\draw[draw=black,fill=dodgerblue52152219,opacity=0.8,thick] (axis cs:0.07,0) rectangle (axis cs:0.17,-89.6408450704225);
\draw[draw=black,fill=dodgerblue52152219,opacity=0.8,thick] (axis cs:0.57,0) rectangle (axis cs:0.67,-89.4513888888889);

\path [draw=black, semithick]
(axis cs:0.12,-89.1448819441368)
--(axis cs:0.12,-90.1368081967082);

\path [draw=black, semithick]
(axis cs:0.62,-86.3452707189749)
--(axis cs:0.62,-92.5575070588029);

\addplot [semithick, black, mark=-, mark size=5, mark options={solid}, only marks]
table {%
0.12 -89.1448819441368
0.62 -86.3452707189749
};
\addplot [semithick, black, mark=-, mark size=5, mark options={solid}, only marks]
table {%
0.12 -90.1368081967082
0.62 -92.5575070588029
};

\draw (axis cs:0.12,-72.1368081967082) node[
  scale=0.6,
  anchor=north,
  text=black,
  rotate=0.0
]{\bfseries -89.6};
\draw (axis cs:0.62,-94.5575070588029) node[
  scale=0.6,
  anchor=north,
  text=black,
  rotate=0.0
]{\bfseries -89.5};

\end{axis}

\end{tikzpicture}
            \vspace{-.18cm}
            \caption{System performance with when applying the optimal configuration identified by \framework.}
            \label{fig:changeintro}
\end{subfigure}
    \caption{Illustrative example that shows the need for configuration parameter tuning. Experimental results are collected using a private 5G network and the framework discussed later in this paper.}
    \label{fig:example}
    \vspace{-.3cm}
\end{figure}
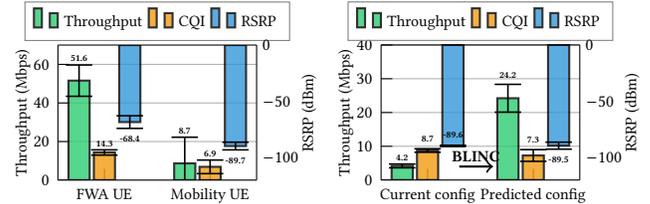

\textbf{Motivation and Example.} Figure \ref{fig:example} demonstrates the necessity of tuning configuration parameters to accommodate deployment-specific conditions.
Consider a network operator that deploys the same \gls{gnb} stack in two different operational environments. One features \gls{fwa} \glspl{ue}, and the other high-mobility \glspl{ue} in vehicular scenarios. 
 
Figure~\ref{fig:baseintro} shows the mean and variation of throughput, \gls{rsrp} and \gls{cqi} for two \gls{ue} types under an \emph{identical} base station configurations in the two deployment scenarios, corresponding to different received power settings.

The \gls{fwa} \glspl{ue} show low throughput variability, reflecting their stable \gls{rf} environment. In contrast, the high-mobility \glspl{ue} experience performance degradation and more throughput variability, with a degraded \gls{rsrp}. 
This indicates that a configuration satisfying the \gls{qos} for the \gls{fwa} scenario does not necessarily guarantee the same in different deployment scenarios.
  
In Fig.~\ref{fig:changeintro}, we focus on the \glspl{kpi} for the high mobility scenario. We measure performance with a default, baseline configuration (left bars, corresponding to a generic, non-deployment-specific set of parameters recommended for the protocol stack we use), and the one optimized using the framework proposed in this paper (right bars). As can be seen, the throughput improves by more than 5 times. This shows the potential of configuration tuning in improving the performance of end-to-end cellular networks.

\textbf{Intuition, Framework Description, and Challenges.} In practice, operators would require technicians with domain expertise to manually tune parameters through iterative testing. This is a time-consuming process. In this work, we propose \framework (Bayesian \gls{llm}-Driven Intelligent Network Configuration). \framework automates network configuration by using Bayesian Networks to learn the joint probability distribution from operational data logs and telemetry, and recommending configurations that improve network performance.

A Bayesian Network is a causal probabilistic graphical model defined by a \gls{dag} with variables as nodes and directed edges between the nodes. The model captures the joint probability distribution associated with each node and its parent set~\cite{rahman1997introduction}.  Bayesian Networks offer several properties that make them suitable for cellular network configuration optimization. \cite{bashar2014application} shows the potential of using Bayesian Networks in binary network management tasks such as admission control and switch port activation.
Unlike rule-based systems that require explicit specification of all parameter interactions or black-box machine learning models, Bayesian Networks represent dependencies as an interpretable directed graph. 
This makes the learned causal relationships between configuration parameters and \glspl{kpi} transparent to network operators. 
Through structure constraints and prior distributions, Bayesian Networks naturally combine domain expertise, addressing the key limitation of purely data-driven methods \cite{marcot2019advances}. 
This methodology also quantifies uncertainty, and enables probabilistic inference for recommendations rather than point predictions and thus aids in efficient representation of uncertainty \cite{barco2002automated}. 
This approach helps in addressing four critical needs: (1) learning complex, non-linear relationships from data without manual feature engineering, (2) identifying configurations that improve performance and avoiding sub-optimal ones through probabilistic reasoning, (3) adapting recommendations to operating context implicitly rather than applying fixed rules, and (4) providing interpretability and uncertainty quantification.

However, pure data-driven structure learning approaches for Bayesian Networks suffer from fundamental limitations. First, statistical ambiguity: multiple DAG structures may fit the data equally well (Markov equivalence), making it impossible to distinguish true causal direction from spurious correlation without additional constraints. Limited data samples in specific operational contexts or a high dimensional parameter space can further lead to unreliable structure estimation.  Second, purely observational data (e.g., based on network telemetry) cannot reveal the causal mechanisms specified in domain knowledge (e.g., protocol specifications) if those relationships are not strongly reflected in the data distribution. This is a limitation of prior work, e.g.,~\cite{bashar2014application}. Structural constraints are usually used to reduce the graph size, but usually need to be provided by domain experts~\cite{de2011efficient}. 
%
Recently, the authors of \cite{ban2025integrating} studied the use of \glspl{llm} to provide reasoning ability in causal analysis, particularly in Bayesian structural learning. 
This motivated us to employ the reasoning ability of \gls{llm} to guide the Bayesian Network construction and perform inference based on it. Compared to the intuition provided in~\cite{ban2025integrating}, we design a framework that captures a wider set of causal relationships, and go beyond structure discovery by leveraging the Bayesian Network to infer optimal configurations for the system.  Crucially, LLM-derived constraints reduce the search space for structure learning algorithms, addressing a fundamental challenge in causal discovery with limited observational data, a benefit that becomes more pronounced as the parameter space grows.
\begin{figure*}[t]
    \centering
    \includegraphics[width=0.6\linewidth]{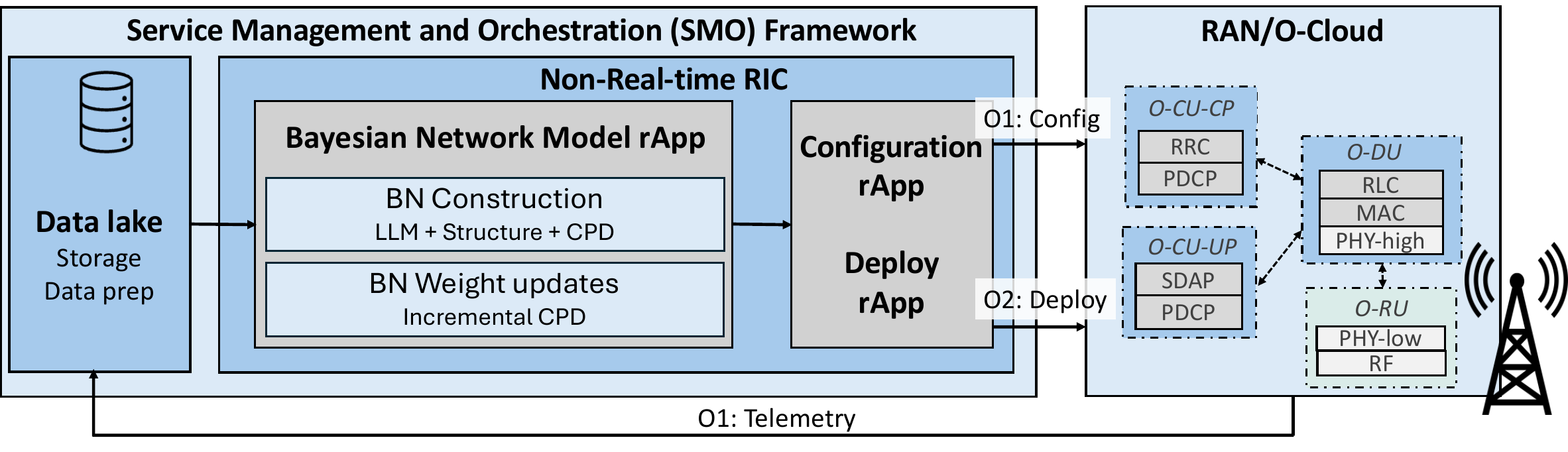}
    \vspace{-.4cm}
    \caption{High-level system diagram with continuous deployment pipeline.}
    \label{fig:update}
    \vspace{-.15cm}
\end{figure*}

\textbf{Our contributions.} In this paper, we address the challenge of automatically optimizing configuration parameters in 5G networks to maximize network performance by designing, prototyping, and evaluating \framework. The main contributions of this paper are as follows:

\begin{itemize}
    \item We propose a domain-constrained structural learning strategy for Bayesian Network construction. It uses an \gls{llm}-enhanced scoring function that combines statistical evidence with causal constraints extracted from 5G domain knowledge, reducing structural errors compared to state-of-the-art data-driven approaches. The Bayesian inference framework can be used for joint multi-parameter optimization through Bayesian inference. Compared to other configuration tuning strategies, our method provides interpretable causal structure and quantifies uncertainty of prediction through confidence scores, enabling operators to assess recommendation reliability before deployment.
    
    \item We integrate the framework, \framework, with the Open \gls{ran} architecture. We design a closed-loop system that collects performance data during automated testing cycles, incrementally updates the Bayesian Network while maintaining the causal structure and adapting only conditional probabilities, and applies optimized configurations in subsequent test iterations. This allows \framework to dynamically adapt to different conditions, which may vary over time or across deployments.
        
    \item We collect a dataset from a private 5G network with diverse RF conditions, capturing various user scenarios  across a controlled sweep of uplink power control configurations and \glspl{kpi} logged during over-the-air testing. To our knowledge, this is the first publicly available dataset linking 5G RAN configuration parameters to measured performance in a real deployment, enabling reproducible research on configuration optimization.\footnote{The dataset will be released upon paper acceptance.} 

    \item We prototype the system and profile it by using different \glspl{llm} and update techniques. We show how the proposed structural learning approach, based on \glspl{llm}-enhanced scoring functions, outperforms other state-of-the-art data-driven techniques. We also show how recent open-source \glspl{llm} perform as well as state-of-the-art proprietary models. This is key as network operator may prefer to run the \gls{llm} scoring in-house rather than exposing telemetry to cloud \gls{llm} providers.
    
    \item We validate \framework in multiple private 5G deployment contexts (i.e., in networks with different ratios of cell-edge and cell-center users), showing up to 63.5\% throughput improvement along with \gls{bler} reduction over default configurations, with \framework outperforming heuristics and single-parameter optimization baselines. 
\end{itemize}

\section{Related Work}
The problem of finetuning various configuration parameter that aims to improve performance of cellular networks has been studied before. Authors of \cite{mahimkar2021auric,mahimkar2022aurora} use statistical association (Chi-square tests) to identify which network attributes like frequency, morphology, traffic  correlate with configuration parameters, and subsequently recommend values through majority voting among carriers with matching attributes. The proposed systems operates based on the assumption that widely deployed configuration settings indicates operational stability and lead to better performance to the users over explicit performance optimization.
Authors of \cite{ge2023chroma}  addresses configuration optimization in cellular networks by learning network contexts where specific configuration changes historically improved performance. 
The system uses supervised classification techniques to identify which network attributes correlate with successful configuration 
changes. 
Another data-driven work for parameter tuning is \cite{ge2025iridescence}. This work aims at handling confounded data and is an extension of \cite{ge2023chroma}. The authors introduces a  de-confounding mechanism that re-labels these samples using a two-stage approach. This enables the proposed method to preserve data and achieve more qualified recommendations.   However, these papers share the same fundamental limitation: configuration recommendation is treated as classification problem without explicitly modeling causal relationships between configurations and performance metrics. In contrast, our approach leverages Bayesian Networks that capture probabilistic causal dependencies between configuration parameters and performance outcomes. Furthermore, some papers, e.g.,~\cite{ge2023chroma}, rely on human domain experts for final validation of recommendations. In contrast, we employ \gls{llm}-assisted causal constraints to explicitly model parameter-performance dependencies with uncertainty quantification and automatically validate the performance with real-world experiments. 

The authors of \cite{patel2024cipat} recognize the importance of intermediate metrics and propose another pure data-driven method that leverages observable intermediate network metrics and predicts the impact of configuration changes. However, the paper focuses on dataset-based analysis. In this work, we close the loop by evaluating the impact of reconfigurations on a real-world private 5G network. \cite{sharma2025towards} proposed a framework for xApp conflict detection in \gls{oran} that combines explainable machine learning (SHAP) with causal inference methods. While their work showcases the importance of causal effect in \gls{ran} management, it does not address configuration optimization and recommendation of optimal parameter settings, as we do, but rather focuses on multi-agent conflict resolution. 
Further, our causal structure learning incorporates domain knowledge through LLM-integration. In contrast, inferring causal structure solely from SHAP scores may produce spurious edges. Also, we provide configuration recommendations with quantified prediction uncertainty.

Machine learning has also been extensively applied to configuration tuning across other domains. These include Bayesian optimization for cloud systems~\cite{alipourfard2017cherrypick, li2018metis}, for network protocols~\cite{bin2024config}, and
multi-armed bandits for Content Distribution Networks~\cite{naseer2017configtron, naseer2022configanator}. These works primarily focus on optimization without explicitly modeling causal relationships between configuration parameters and performance outcomes. While machine learning has been widely applied to network optimization, our approach offers advantage on generalization as the causal structure can transfer across environments easily. In contrast, machine learning policies including \gls{drl} require retraining when conditions change, and lack intrinsic interpretability and uncertainty quantification enabling operators to assess recommendation reliability.

\section{System model}
\label{sec:systemmodel}

We consider a cellular network setting with a 5G base station, known as \gls{gnb}, which provides wireless communication to the users. Following the Open RAN architecture, the \gls{gnb} is disaggregated into modular components, or \gls{ran} functions, with open interfaces: \gls{cu}, \gls{du}, and a \gls{ru} (Fig.~\ref{fig:update}). 
The \gls{oran} architecture also introduces \glspl{ric}, which enable \gls{ran} optimization and automation with closed-loop control via different types of applications. rApps are used in the \gls{nrtric}, xApps in \gls{nrric}, and dApps on the \gls{ran} functions. Further, a \gls{smo} coordinates service deployment, configuration, and monitoring, in coordination with the \gls{nrric}. The architecture enables to capture and expose \glspl{kpi} through open interfaces~\cite{bonati2021intelligence}.

Modern 5G base stations manage radio resource allocation, link adaptation, power control, and scheduling decisions. Their configuration includes several tunable parameters that span the whole protocol stacks. For example, these parameters include physical layer settings like antenna configuration, power control parameters like maximum transmit power, \gls{rach} configuration parameter like power ramping. The parameters jointly influence the network performance through their interactions with channel and \gls{rf} conditions. Given this parameter configuration space, we introduce a causal modeling framework for predicting optimal \gls{ran} configurations that maximize network performance. Here, \textit{causal} 
means that an edge $A \rightarrow B$ indicates actively changing $A$ 
produces a measurable change in $B$. 
This is distinct from statistical 
association, where $A$ and $B$ merely co-vary in observational data.
This distinction is critical for configuration recommendation as acting on a variable that is an effect rather than a cause produces no performance improvement. Our framework combines domain knowledge with statistical evidence: the \gls{llm} constrains the hypothesis space, and the structure learning algorithm selects among feasible structures from data, yielding a \gls{dag} that reflects both prior knowledge and empirical observation.  

From a system perspective, the Bayesian Network framework along with the configuration and deployment components are designed as rApps deployed on the \gls{nrtric}. The rApps perform model construction and updates, using network telemetry, and generate configuration recommendations applied via CI/CD pipelines. Figure~\ref{fig:update} illustrates the closed-loop system architecture with four processes: \Circled{1} data collection from network operation, \Circled{2} model construction or update using accumulated data, \Circled{3} configuration generation via Bayesian inference and deployment, and \Circled{4} network operation with applied configurations, completing the feedback loop.

\subsection{Bayesian Network Construction and Configuration Recommendation}
 This section discusses the Bayesian Network construction and recommendation strategies. The system operates in two main phases: \gls{llm}-assisted Bayesian Network construction and Bayesian Network-guided configuration generation with multi-parameter optimization.

\begin{figure}[!h]
\vspace{-.2cm}
    \centering
    \includegraphics[width=\linewidth]{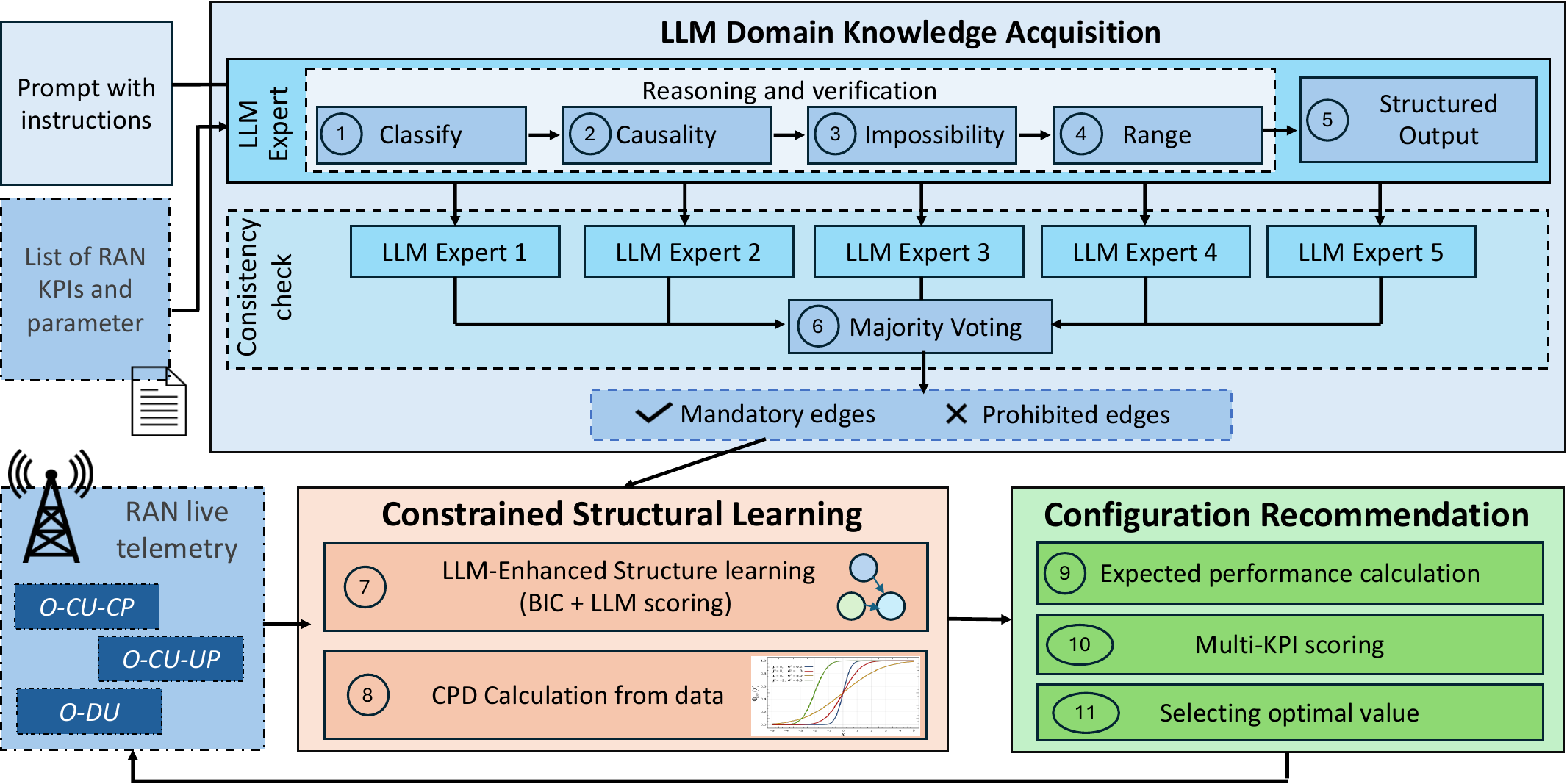}
    \vspace{-.4cm}
    \caption{Workflow and components involved in the Bayesian Network construction.}
    \vspace{-.25cm}
    \label{fig:system}
\end{figure}

Figure~\ref{fig:system} presents the high-level workflow and algorithm architecture, comprising three major components: \gls{llm} domain knowledge acquisition, constrained structure learning, and configuration recommendation. The constrained structure learning component integrates processed network data with domain expertise extracted from \gls{llm}-based domain experts. The resulting learned structure, together with estimated \glspl{cpd}, enables configuration recommendations for network optimization.

\subsubsection{Data Preparation}
The raw network \glspl{kpi} and  parameter values are continuous in nature.
These continuous values needs to be discretized while preserving the underlying statistical relationships to make it compatible with the bayesian learning algorithms and associated conditional probability table construction. We transform the continuous variables into categorical states using quantile-based binning. The quantile-based approach ensures balanced representation across performance ranges and maintains ordinal relationships between categories. This preprocessing step is essential because Bayesian Network inference algorithms require discrete variables for tractable conditional probability distribution computation and ensures robust statistical estimation from finite experimental datasets. We note that configurable parameters  retain their original discrete values since these represent specific design choices rather than measured quantities.
The data preparation also includes a prompt engineering step, which provides the \gls{llm} model with specific instructions and context. 
This includes information such as the details of general operational context, parameter classification guide, reasoning approach.

\subsubsection{LLM Domain Knowledge Acquisition}
\begin{figure}
    \centering
    \begin{subfigure}{0.48\linewidth}
        \includegraphics[width=\linewidth]{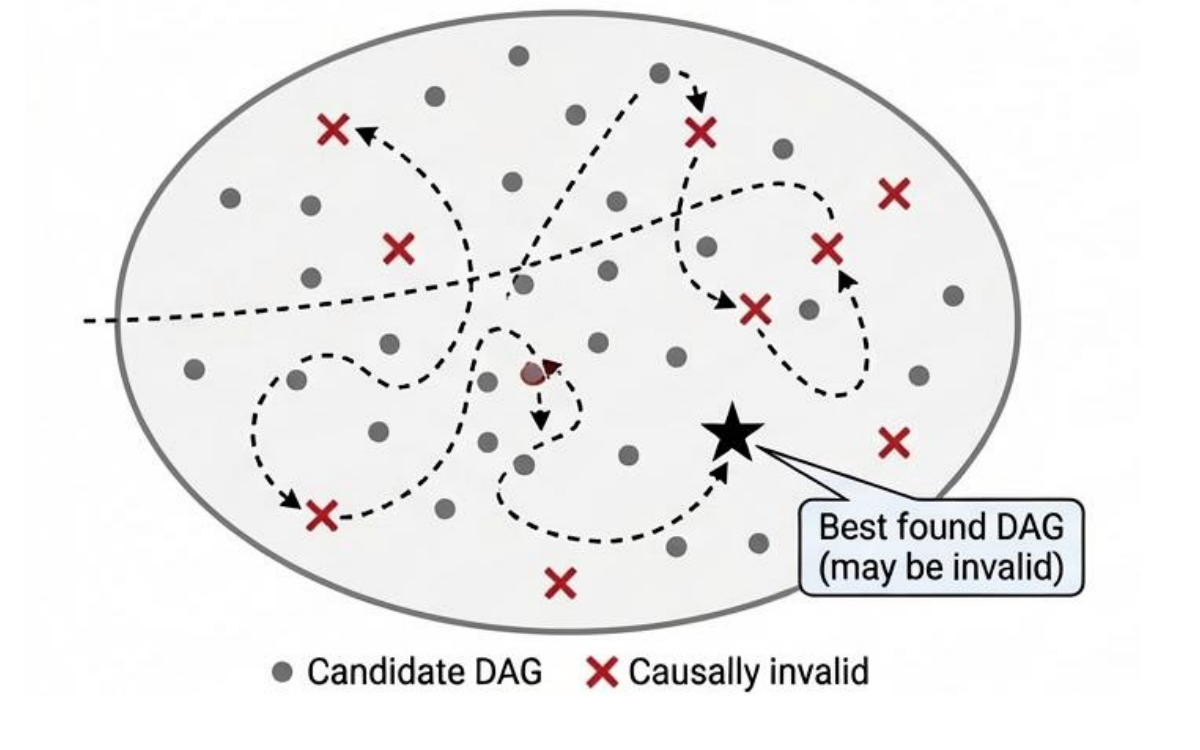}
        \caption{Unconstrained search space}
        \label{fig:un_search}
    \end{subfigure}
    \hfill
    \begin{subfigure}{0.48\linewidth}
        \includegraphics[width=0.9\linewidth]{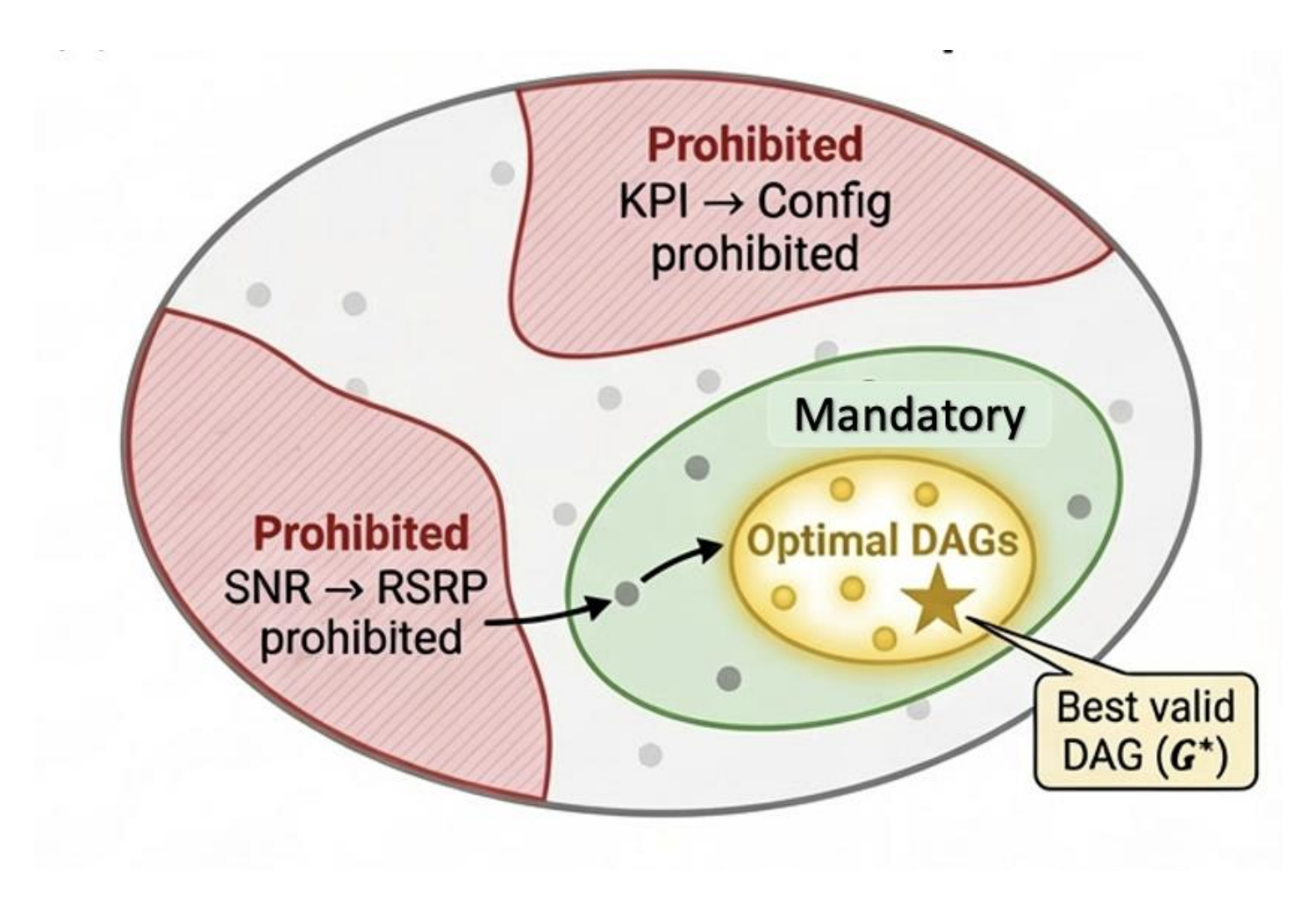}
        \caption{LLM constrained search space}
        \label{fig:llm_search}
    \end{subfigure}
    \caption{Illustration of unconstrained and \gls{llm}-constrained search spaces for Bayesian Network structure learning.}
    \vspace{-.2cm}
\end{figure}

The first step in Bayesian Network construction is construction of \gls{dag} which represents the causal structure among variables.  However, discovering domain specific causal structures from observational data is a challenging task~\cite{de2011efficient}. Integrating domain knowledge to the structure learning framework is known to reduce the search space and enhance the accuracy of the structure learned~\cite{constantinou2023impact}. 
Without domain constraints, the structure learning algorithm explores an unconstrained search space, and may converge to invalid DAG structures containing spurious or reversed edges (Fig.~\ref{fig:un_search}). By restricting the search space by prohibiting edges that violate physical or protocol constraints and enforcing edges that reflect known causal mechanisms, we guide the algorithm toward feasible structures from which a correct DAG can be recovered (Fig.~\ref{fig:llm_search}). 
Recently, the reasoning capabilities of \glspl{llm} have been explored for tasks involving causal discovery and analysis~\cite{zhang2023understanding,ban2025integrating,ban2025llm}.
Thus, in this work, we leverage \gls{llm} to integrate domain knowledge to the structure learning framework and aim to increase accuracy of the Bayesian Network learned. 

To achieve this goal, we feed \gls{llm} model with a human-designed detailed context and instructions along with the list of \glspl{kpi} observed by the \gls{gnb} and parameter variables that are configured. The prompt is designed once and then used consistently across all experiments. It consists of three stages: reasoning and verification, structured output, and consistency check. The reasoning and verification stage maps the variables to causality. In this stage, the \gls{llm} follows a four-step chain-of-thought process, where each step builds on prior reasoning, mirroring the systematic approach a domain expert would take when analyzing parameter relationships from first principles. These steps are as follows:
\Circled{1} Parameter classification stage: In this stage LLM is prompted to classify the variables as control, measurements, performance metric types. \Circled{2} Causality tracing stage: This stage obtains information on which parameters influence others along with reason and results in potential edges between the variables. \Circled{3} Checking impossibility conditions: In this next step, relationships that violate temporal/physical/protocol constraints are checked. \Circled{4} Finally, ranges of each parameters are also gathered. 
A fifth step verifies the causality results. The instructions prompt the \gls{llm} to verity that the relation is based on 3GPP specification, and should have explanation on the direction of causality and the mechanism involved. Thus at this stage, i.e., at the end of \Circled{4}, the \gls{llm} expert provides a detailed justification for each causal relationship it identifies.
%
The Structured Output stage presents the causality learned as:
\begin{itemize}
    \item Mandatory edges: Causal relationships that must exist based on physical mechanisms, temporal ordering, or protocol specifications;
    \item Prohibited edges: Relationships that violate causality, including reverse temporal ordering (effect→cause), physically impossible mechanisms, and protocol violations.
\end{itemize}  

To improve robustness against \gls{llm} variability, we adopt an ensemble strategy, executing the full chain-of-thought reasoning process $n$ times independently and aggregating results through majority voting  retaining only constraints that appear in more than 50\% of responses.
An example of the causal relationships learned is presented in Tab.~\ref{tab:example_constraints}. 
We note that manual constraint engineering does not scale as a replacement for the \gls{llm} as enumerating pairwise causal relationships among $|\mathcal{V}|$ variables requires evaluating $O(|\mathcal{V}|^2)$ candidate edges, each demanding reasoning about physical mechanisms and cross-layer interactions, an effort that must be repeated whenever the variable set changes.  

\begin{table}[t]
\centering
\caption{Example of mandatory and prohibited causal relationships obtained from the LLM.}
\vspace{-.3cm}
\footnotesize
\begin{tabular}{lllp{4cm}}
\toprule
Type & Source & Target & Reasoning \\
\midrule
MANDATORY & p0\_nominal & RSRP & p0\_nominal controls UE uplink transmit power according to 3GPP 38.213, affecting received signal power at gNB \\
MANDATORY & RSRP & SNR & RSRP is a direct component in SNR calculation; higher received power improves signal quality relative to noise \\
PROHIBITED & UL\_Mbps & p0\_nominal & Performance metrics cannot directly modify control parameters without operator intervention \\
PROHIBITED & UL\_BLER & p0\_nominal & Error rates are outcomes of configurations, not direct inputs to power control settings \\
\bottomrule
\end{tabular}
\label{tab:example_constraints}
\vspace{-.6cm}
\end{table}

\subsubsection{Constrained Structure Learning}
\label{sec:structure}
We then leverage the information on causal relationships and construct a Bayesian Network to model causal dependencies among configuration parameters and \glspl{kpi} based on \gls{ran} telemetry. The structure of the Bayesian Network is represented as a \gls{dag} $\mathcal{G} = (\mathcal{V}, \mathcal{E})$, where nodes $\mathcal{V}=\{v_1, v_2, \ldots, v_n\}$ represent network variables which includes configuration 
parameters and \glspl{kpi}, and directed edges $\mathcal{E}$ capture causal dependencies. 
Let the parents of each node $v_i$ is represented as $\text{Pa}(v_i)$. To find the optimal parameter combinations using probabilistic inference, we require the joint probability distribution over all variables. The joint probability distribution of the Bayesian Network is defined as follows:
\begin{equation}
    P(v_1,v_2..v_n)=\prod\limits_{i \in V} P(v_i | \text{Pa}(v_i)).
\end{equation}
The structural learning problem learns the \gls{dag} $\mathcal{G}$ that represents causal relationships from data $D$.  Traditional structure learning algorithms for \glspl{bn} leverages statistical scoring functions like \gls{bic}~\cite{suzuki1996learning}, K2~\cite{cooper1992bayesian} or \gls{bdeu}~\cite{buntine1991theory}. \gls{bic} evaluates the \gls{dag} structures by using a complexity penalty term along with the data log-likelihood score and tries to balance model fit accuracy with less complexity. K2 evaluates the candidate structures by computing posterior probability of each  \gls{dag} given the data using Bayesian approach. \gls{bdeu} works in a similar way but uses prior based on equivalent sample size. To integrate domain knowledge constraints with data-driven scoring,  we use an \gls{llm}-enhanced scoring function. Let the dataset used be denoted as $D$ and the constraints (mandatory and prohibited edges learned using the LLM) as $\Delta=\{\Delta_m,\Delta_p\}$. The scoring function evaluates the candidate graph structures using the dataset $D$ and constraints $\Delta$.
The scoring function is defined as follows:
\begin{equation}
   S(\mathcal{G},D,\Delta)=S_{stat}(\mathcal{G},D) + S_{\gls{llm}}(\mathcal{G},\Delta). 
\end{equation}
Here, $S_{stat}(\mathcal{G},D)$ represents base score learned from data using any statistical scoring function like \gls{bic} or K2 scores. $S_{\gls{llm}}(G,\Delta)$ represents the additional score that is added for mandatory and prohibited edge constraints learned by the \gls{llm}. This is defined as follows: 
\begin{align}
    S_{\gls{llm}}(\mathcal{G},\Delta) &= S_{\gls{llm}}(\mathcal{G},\Delta_m) + S_{\gls{llm}}(\mathcal{G},\Delta_p)\\
                 = &\sum_{{(u,v) \in \Delta_m}} [\phi(u,v)  \cdot \alpha_{reward} - (1-\phi(u,v))  \cdot \alpha_{penalty}] \nonumber \\
                 &\hspace{10mm}- \sum_{{(u,v) \in \Delta_p}} \beta_{penalty}.
\end{align}
The score has two components for mandatory and prohibited edge constraints, denoted as $S_{\gls{llm}}(\mathcal{G},\Delta_m)$ and  $S_{\gls{llm},p}(\mathcal{G},\Delta_p)$ respectively. In the mandatory edge component, a reward $\alpha_{reward}$ is applied when the edge is present in the candidate graph $\mathcal{G}$ where as $\alpha_{penalty}$ when it is not present.
Here, $\phi(u,v)$ is an indicator function that denote whether the edge is present in $\mathcal{G}$ or not and is defined as
\begin{equation}
\phi(u, v) = 
\begin{cases}
1 & \text{if } (u,v) \in \mathcal{G}, \\
0 & \text{otherwise}.
\end{cases}
\end{equation}
At the same time, the prohibited edge component penalizes candidate graphs containing edges that violate domain constraints. When a prohibited edge $(u,v) \in P$ is present in candidate graph, a penalty score $\beta_{penalty}$ is applied. We note that the approach uses hard constraints that ensures that the edges that are mandatory must be present in the graph and prohibited are not. 

We employ the hill-climbing algorithm \cite{selman2006hill,koller2009probabilistic} to search the space of possible DAG structures, using the \gls{llm}-enhanced scoring to evaluate candidate 
structures at each step. The algorithm iteratively modifies the edges to maximize the combined score and maintaining the acyclic nature of the structure.
%
The \gls{dag} structure learned in the previous step along with the statistical data is used to generate the complete Bayesian Network. The process involves estimating the network parameters by computing \glspl{cpd} for each node in $\mathcal{G}$ from the available data.

\subsubsection{Configuration Generation}
After construction of the Bayesian Network $\mathcal{G}$, the conditional probabilities learned can be used to for probabilistic inference and configuration optimization. 
Let $\boldsymbol{\Theta} = \{\theta_1, \theta_2, \allowbreak \ldots, \theta_r\}$ denote the set of \glspl{cpd} associated with $\mathcal{G}$,  where each  $\theta_i$  represents  $P(v_i \mid \text{Pa}(v_i))$. 

We partition the variables $\mathcal{V}$ into three disjoint sets:
\begin{itemize}
    \item $\mathcal{C} = \{c_1, c_2, \ldots, c_k\}$: configurable parameters (e.g., p0\_nominal, pusch\_dtx\_threshold)
    \item $\mathcal{M} = \{m_1, m_2, \ldots, m_l\}$: measurement variables (e.g., SNR, RSRP)
    \item $\mathcal{K} = \{k_1, k_2, \ldots, k_q\}$: target KPIs (e.g., UL\_Mbps, UL\_BLER)
\end{itemize}
where $|\mathcal{C}| = k$, $|\mathcal{M}| = l$, $|\mathcal{K}| = q$, and 
$\mathcal{V} = \mathcal{C} \cup \mathcal{M} \cup \mathcal{K}$.
Given the current configuration $\mathbf{c}{(t)} = (c_1{(t)}, c_2{(t)}, \ldots, c_k{(t)})$ and current measurements $\mathbf{m}{(t)} = (m_1{(t)}, m_2{(t)}, \ldots, m_l{(t)})$, the objective is to find the optimal configuration $\mathbf{c}^*$ that maximizes the expected performance:
\begin{equation}
    \mathbf{c}^* = \arg\max_{\mathbf{c} \in \Omega} \sum_{i=1}^q w_i \cdot \mathbb{E}[f_i(k_i) \mid \mathbf{c}, \mathbf{m}{(t)}],
    \label{eq:objective}
\end{equation}
where $\Omega$ is the feasible configuration space, $w_i$ is importance weights for each \gls{kpi} $k_i \in \mathcal{K}$, and $f_i(\cdot)$ are utility functions mapping \gls{kpi} values to performance scores with positive values for benefit metrics like throughput and negative values for cost metrics like error rates. The optimal configuration generation is detailed in Algorithm~\ref{algo:conf}. The algorithm takes the learned Bayesian Network as input. 
For each candidate configuration $\mathbf{c}\in \Omega$, we compute the posterior distribution of the target KPIs using the learned conditional probability distributions of Bayesian Network as follows:
\begin{align}
P(k_i = k \mid \mathbf{c}, \mathbf{m}(t)) = \frac{P(k_i = k, \mathbf{c}, \mathbf{m}(t))}{P(\mathbf{c}, \mathbf{m}(t))}.
\end{align}
With the computed posterior distributions, the expected utility for each KPI is then computed as:
\begin{equation}
\mathbb{E}[f_i(k_i) \mid \mathbf{c}, \mathbf{m}{(t)}] = \sum_{s \in \mathcal{S}_{k_i}} f_i(s) \cdot P(k_i = s \mid \mathbf{c}, \mathbf{m}{(t)}),
\end{equation}
where $\mathcal{S}_{k_i}$ denotes the set of possible discrete states for KPI $k_i$.
The algorithm then searches and select configuration that maximizes the sum of expected utilities of the target KPIs as the recommended configuration.

\begin{algorithm}[t]
\caption{Bayesian Network-Based Configuration Generation.}
\footnotesize
\label{algo:conf}
\begin{algorithmic}[1]
 \renewcommand{\algorithmicrequire}{\textbf{Input:}}
     \renewcommand{\algorithmicensure}{\textbf{Output:}}
\REQUIRE Bayesian Network $\mathcal{G}$,  weights $\mathbf{w}$
\ENSURE Recommended configuration $\mathbf{c}^*$
\STATE Initialize $S_{\max} = -\infty$, $\mathbf{c}^* = \mathbf{c}(t)$
\FOR{each candidate configuration $\mathbf{c} \in \Omega$}
    \FOR{each target KPI $k_i \in \mathcal{K}$}
        \STATE Compute $P(k_i \mid \mathbf{c}, \mathbf{m}(t))$ 
        \STATE Calculate $\mathbb{E}[f_i(k_i) \mid \mathbf{c}, \mathbf{m}{(t)}] = \sum_{s \in \mathcal{S}_{k_i}} f_i(s) \cdot P(k_i = s \mid \mathbf{c}, \mathbf{m}{(t)})$
        \STATE Update $S(\mathbf{c}) \leftarrow S(\mathbf{c}) + w_i \cdot \mathbb{E}[f_i(k_i) \mid \mathbf{c}, \mathbf{m}{(t)}]$
    \ENDFOR
    \IF{$S(\mathbf{c}) > S_{\max}$}
        \STATE $S_{\max} = S(\mathbf{c})$, $\mathbf{c}^* = \mathbf{c}$
    \ENDIF
\ENDFOR
\RETURN $\mathbf{c}^*$
\end{algorithmic}
\end{algorithm}







\subsection{Continuous Model Adaptation}



In this section, we present continuous model adaptation strategies as network evolve over time. This is critical to make sure that the network maintains an optimal configuration even with changes and updates in its operational conditions. 

Continuous adaptation over time involves model updates in short and large time scales.
At short time intervals, we incrementally update the weight of the Bayesian Network learned as presented in BN weight updates part of Bayesian Network model rApp in Fig.~\ref{fig:update}. The incremental \gls{cpd} weight update algorithm is detailed in Algo. \ref{algo:update}. It takes the existing Bayesian Network ($\mathcal{G}$ with \glspl{cpd} $\Theta$), the new data collected, $\mathcal{D}_{new}$, and learning rate, $\gamma$ as input. For each node in the Bayesian Network, the algorithm  finds the \gls{cpd} estimates from the new data collected ($\theta_{new}^{(i)}, \forall i \in \{1 \cdots n\}$). This is used to update the old \gls{cpd} incrementally using the learning rate $\gamma$ as follows: 
\begin{equation}
    \theta_{updated}^{(i)} = (1-\gamma) \cdot \theta^{(i)} + \gamma \cdot \theta_{new}^{(i)}.
\end{equation}

The updated \glspl{cpd} are then normalized and used to update the Bayesian Network. By performing periodic finetuning, the model captures short fluctuations over time.
At a larger time scale, we have the option for full learning with all the data collected to maintain integrity of the learned model. This can also be performed with the new data collected if there are any major changes in the network. After finetuning, the learned Bayesian Network uses Algo. \ref{algo:conf} to generate the optimal configuration recommendations, which translate into configuration updates submitted to the \gls{ran} through the O1 interface, or into new deployments using the O2 interface.
\begin{algorithm}[t]
\setlength\abovecaptionskip{-.6cm}
\caption{Incremental \gls{cpd} Weight Update.}
\label{algo:update}
\footnotesize
\begin{algorithmic}[1]
 \renewcommand{\algorithmicrequire}{\textbf{Input:}}
     \renewcommand{\algorithmicensure}{\textbf{Output:}}
\REQUIRE Existing BN $\mathcal{G}$ with \glspl{cpd} $\Theta$, new data $\mathcal{D}_{new}$, learning rate $\gamma$
\ENSURE Updated BN $\mathcal{G}_{updated}$ with \glspl{cpd} $\Theta_{updated}$
\STATE Initialize $\mathcal{G}_{updated} \leftarrow \mathcal{G}$ 
\FOR{each node $v_i \in \mathcal{G}$}
    \STATE Extract old \gls{cpd}: $\theta^{(i)} = P(v_i \mid \text{Pa}(v_i))$
    \STATE Estimate new \gls{cpd} from $\mathcal{D}_{new}$: $\theta_{new}^{(i)} = P_{new}(v_i \mid \text{Pa}(v_i))$
    \IF{$\text{shape}(\theta^{(i)}) = \text{shape}(\theta_{new}^{(i)})$}
        \STATE Update: $\theta_{updated}^{(i)} = (1-\gamma) \cdot \theta^{(i)} + \gamma \cdot \theta_{new}^{(i)}$
        \STATE Normalize: $\theta_{updated}^{(i)} = \frac{\theta_{updated}^{(i)}}{\sum \theta_{updated}^{(i)}}$
        \STATE Add to network: $\mathcal{G}_{updated}.\text{add\_cpd}(\theta_{updated}^{(i)})$
    \ELSE
        \STATE Keep original: $\mathcal{G}_{updated}.\text{add\_cpd}(\theta^{(i)})$
    \ENDIF
\ENDFOR
\RETURN $\mathcal{G}_{updated}$
\end{algorithmic}
\end{algorithm}


\section{Prototyping \framework in an Automated Private 5G Framework}

We integrate the \framework Bayesian Network configuration optimization framework with an existing Open RAN CI/CD/CT infrastructure, 
which automates build, deployment, and testing workflows for the RAN components. This provides a practical implementation of the logical architecture shown in Fig.~\ref{fig:update}. The baseline workflow pulls source code, builds container images for \gls{ran} functions, deploys on the private 5G infrastructure, executes over-the-air testing with \glspl{ue}, and collects performance \glspl{kpi}. All of this is done without manual intervention. With the integration of \framework, we also include a dynamic configuration parameters tuning. The \glspl{kpi} collected and saved during testing are used to construct the Bayesian Network and to perform configuration recommendations. Recommended optimal parameters are applied in the next deployment step of CI/CD iteration. This creates closed-loop learning where each test cycle refines configuration recommendations, achieving continuous performance improvement.


\begin{figure}[!h]
\centering
\vspace{-.3cm}

    \includegraphics[width=0.5\linewidth]{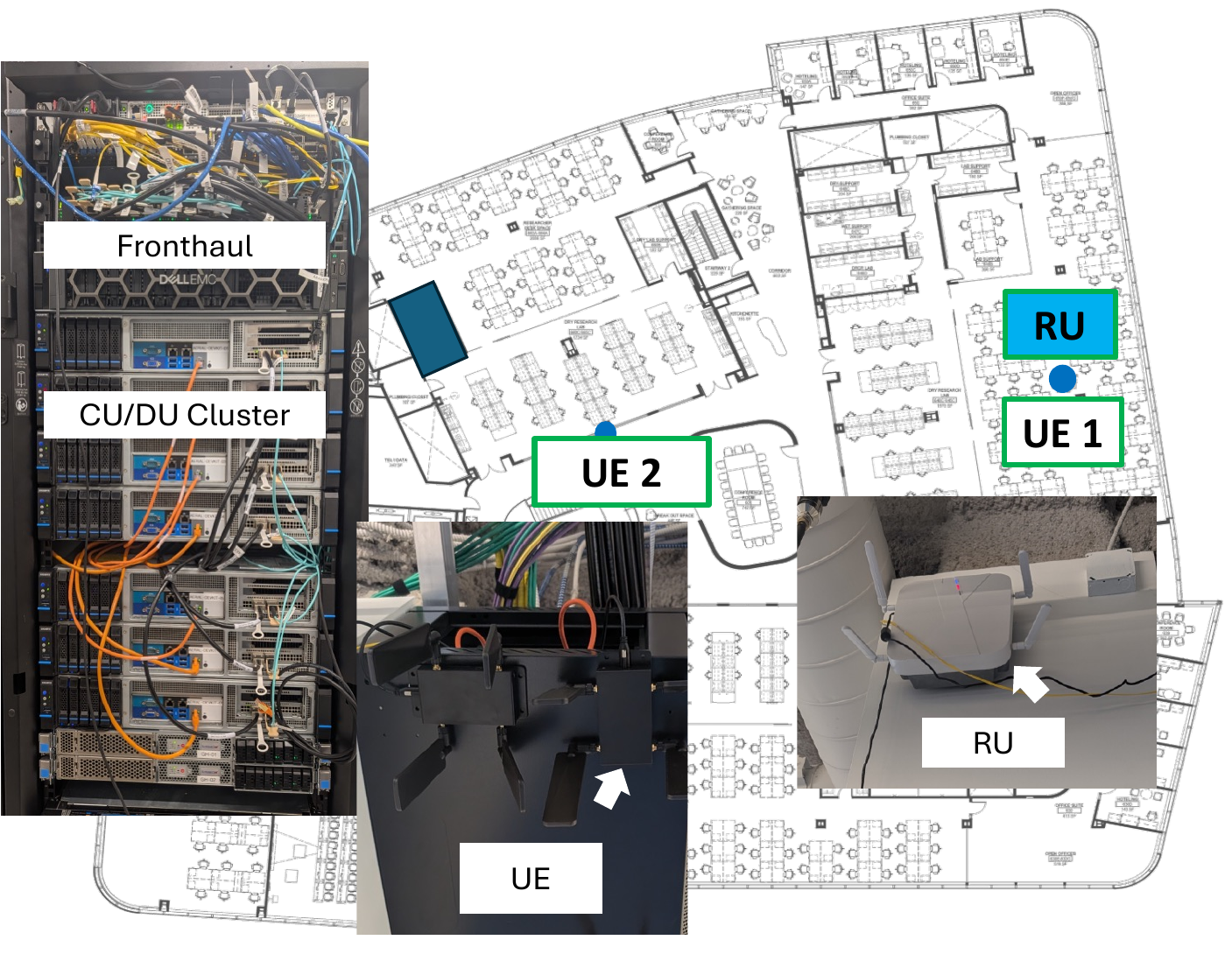}
    \caption{Evaluation setup with locations of \glspl{ru} and \glspl{ue}.}
    \label{fig:evaluation}
    \vspace{-.1cm}

\end{figure}
\begin{figure}
\begin{minipage}{0.48\textwidth}
    \centering
    \footnotesize
    \captionof{table}{Parameter and KPI nomenclature mapping.}
    \vspace{-.3cm}
    \label{tab:nomenclature}
    \begin{tabular}{p{1.8cm}p{2.3cm}p{3.7cm}}
    \toprule
    \textbf{Paper Notation} & \textbf{OAI Parameter} & \textbf{Description} \\
    \midrule
    $P_0$ & \texttt{p0\_nominal} &
    Base transmit power level for PUSCH power control (dBm) \\
    $\Gamma$ & \texttt{pusch\_TargetSNRx10} &
    Target SNR for uplink power control (dB) \\
    UL Throughput & \texttt{UL\_Mbps} & Uplink throughput (Mbps) \\
    BLER & \texttt{UL\_BLER} & Uplink block error rate  \\
    MCS & \texttt{UL\_MCS} & Modulation and coding scheme  \\[0.3em]
    \bottomrule
    \end{tabular}
\end{minipage}
 \vspace{-.35cm}
\end{figure}

The experimental evaluation and data collection are carried out on a 
private 5G network, with PHY layer implemented on \gls{gpu} (i.e., NVIDIA Aerial) with OpenAirInterface (OAI) for the
higher layers of the 5G stack. We use Open5GS as the core network. The \gls{ru} we use is a Foxconn RPQN, configured for 100~MHz bandwidth in band N77 with $4\times4$ MIMO.  Two Sierra Wireless EM9293 modules serve as \glspl{ue}, generating 50 Mbps uplink \gls{udp} traffic via iperf3, representing typical real-time application scenarios. \glspl{ue} are positioned in an indoor space as shown in Fig. \ref{fig:evaluation}. One \gls{ue} is near and directly in \gls{los} of the \gls{ru}. This \gls{ue} has good channel conditions. The other \gls{ue} is placed far in \gls{nlos}. It is located in a corridor with a thick wall separating \gls{ru} and the \gls{ue} creating weak channel conditions.

Table~\ref{tab:nomenclature} provides the mapping between our simplified 
nomenclature for the parameters and target \glspl{kpi} reported and OAI implementation parameter names for clarity.
The configuration space $\Omega$ consists of nominal power  $P_0 \in \{-106, -102, -96, -90, -84\}$ dBm,
and target \gls{snr} $\Gamma \in \{12, 15, 20, 28\}$ dB, yielding 20 unique configurations. As shown in Table~\ref{tab:nomenclature}, $P_0$ is the base transmit power for uplink transmissions, and $\Gamma$ represents the \gls{snr} value that an uplink transmission should target. 
We collected 87,796 measurements across the 20 configurations over 21 days. Each measurement corresponds to the \gls{kpi} values logged by the \gls{gnb} at fixed intervals during each experiment. The dataset consist of the applied configuration parameters and \glspl{kpi} reported. The CI/CD framework applied each configuration through a parametric sweep, ran over-the-air test for a fixed measurement window, and logged L1-reported throughput together with \glspl{kpi} from \gls{oai}. The data set covers both cell-center and cell-edge scenarios. Our 
approach prioritizes repeated measurements to capture \gls{rf} variability.  
\section{Results}
\label{sec:results}

In this section, we discuss the results of our approach. First, we review the impact of the \gls{llm} in structural learning. Second, we evaluate the performance of the configuration optimization with the learned Bayesian Network.

\subsection{\gls{llm} in Structural Learning}
As discussed in the earlier section, data-driven structure learning based only on correlation-driven scoring (e.g., \gls{bic}, K2, BDeu) is challenging in terms of search space and accuracy. The constructed \gls{dag} structure can differ from the ground truth, with missed and reversed edges. In the case of missed edges, the algorithm fails to discover causal relationships. For reversed edges, the algorithm learns correlations in the wrong direction pointing from effect to cause.

\begin{figure}[t]
    \centering
    \vspace{-.4cm}
    \includegraphics[width=0.7\linewidth]{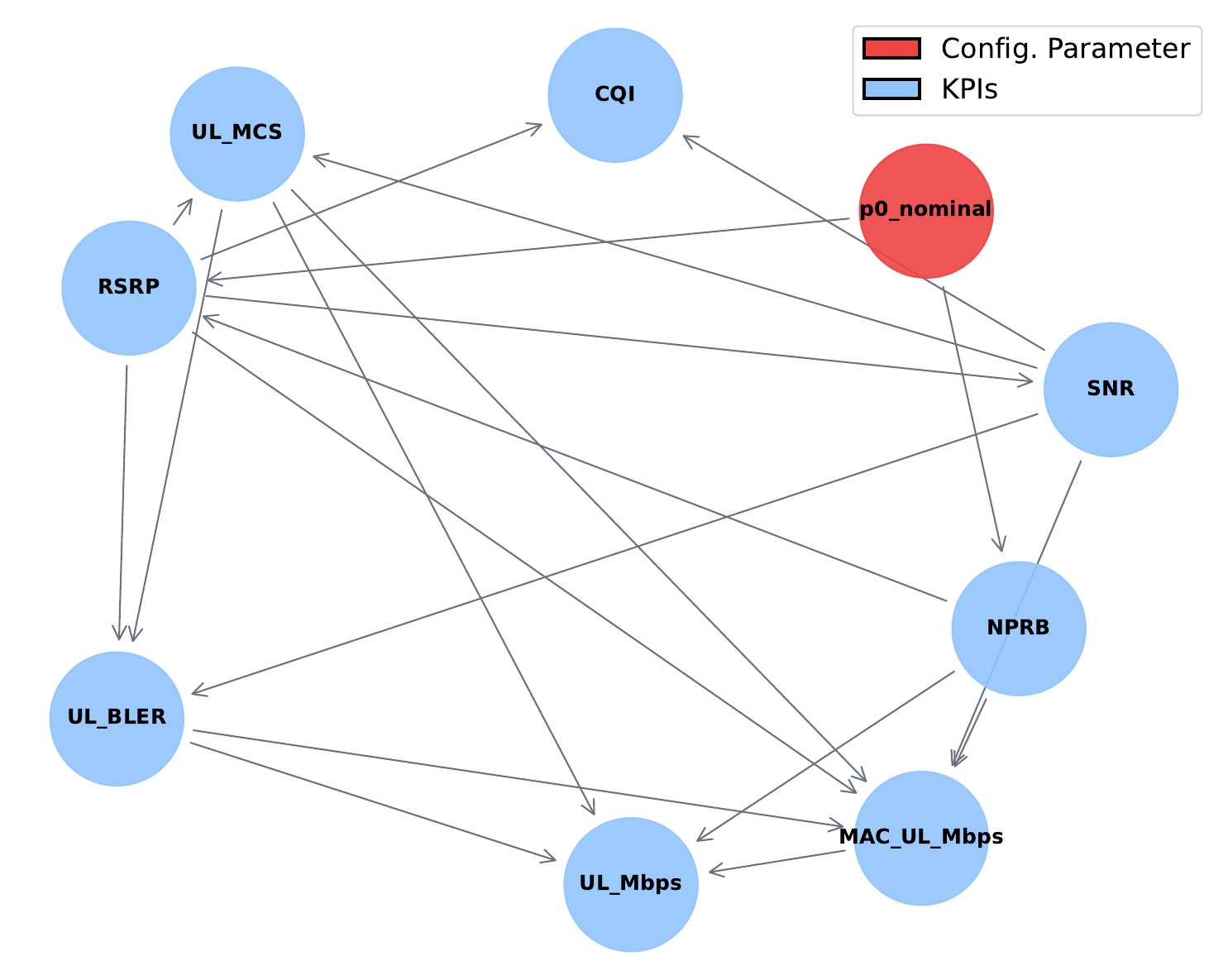}
    \includegraphics[width=0.7\linewidth]{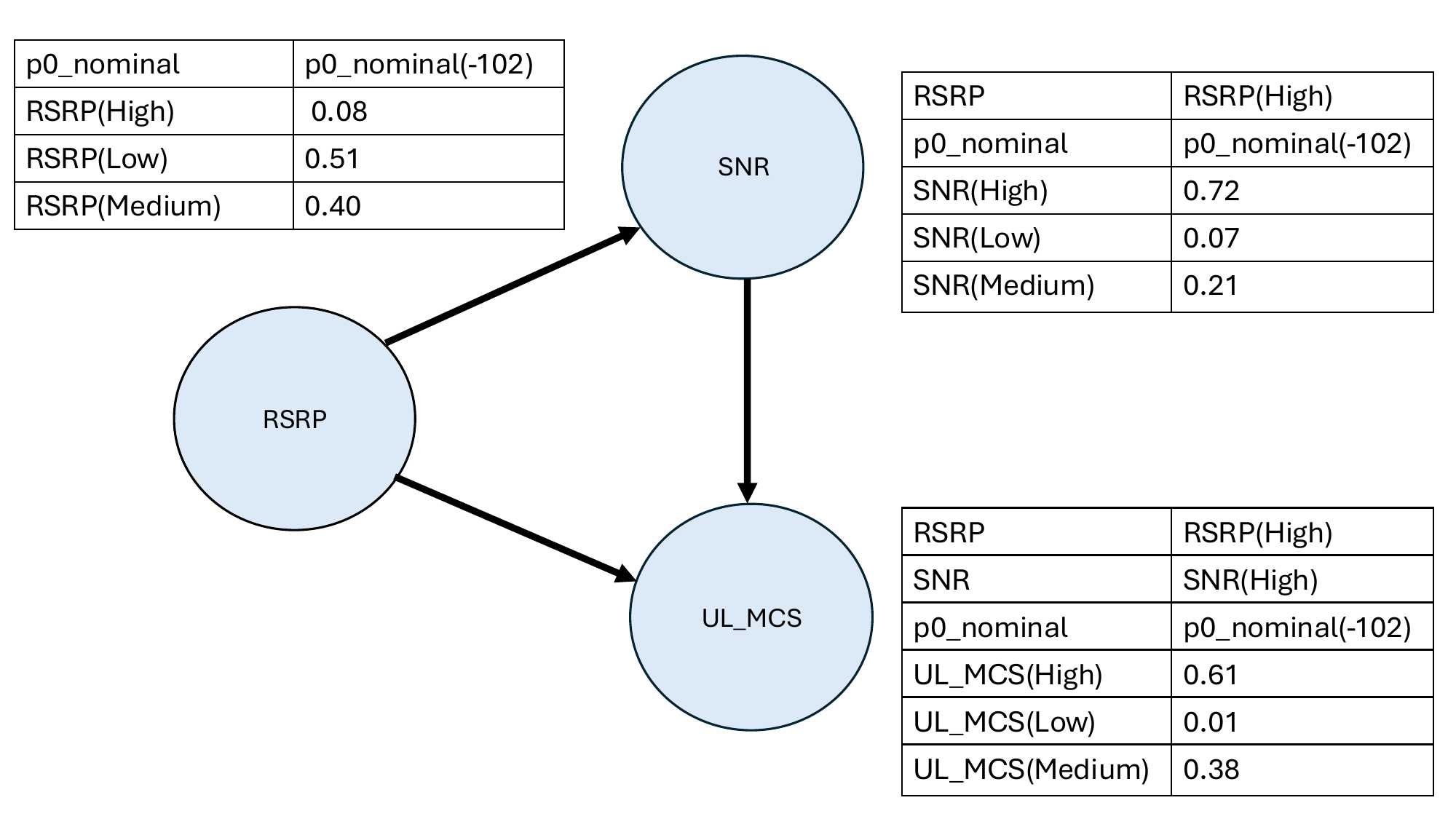}
    \vspace{-.3cm}
    \caption{Example of Bayesian Network learned for parameter $P_0$ and samples from the learned \gls{cpt} associated with nodes.}
    \label{fig:bn}
    \vspace{-0.1cm}
\end{figure}

To study the impact of \glspl{llm} in causal learning, we build the \gls{dag} with and without \gls{llm}, and also using different \glspl{llm}. We compare state-of-the-art commercial models (Claude 4.5 Sonnet) and open-source models (gpt-oss-120b, deployed on an NVIDIA DGX Spark). 
An example of Bayesian Network learned for parameter $P_0$ is presented in Fig.~\ref{fig:bn}. The structure of the Bayesian Network with 
causal dependencies is represented on the left portion of the figure. The right portion of the figure shows a subset of the structure and one entry from the \glspl{cpt} that demonstrate 
how $P_0$ value influences network \glspl{kpi}. 
In this subgraph, the \gls{rsrp} node is a parent for both the \gls{snr} and the uplink \gls{mcs} nodes, while the \gls{snr} node is a parent for the uplink \gls{mcs} node. 
In this example, the \gls{cpt} of the \gls{rsrp} node shows that when the parameter $P_0$ is $-102$ dBm, the \gls{rsrp} is most likely to fall in medium or low category, with probabilities $0.51$ and $0.40$, respectively.  Continuing this inference chain from $P_0$ and the \gls{rsrp} to the uplink \gls{mcs}, the framework captures conditional probabilities in each step conditioned on parents enabling inference queries in the causal path. For example, the \gls{cpt} displayed in the uplink MCS shows the probability of having a high, medium, or low \gls{mcs} when SNR is high, RSRP is high, and $P_0=-102$ dBm.
Similarly, the Bayesian Network captures the probability distribution of all possible parent-child value combinations at each  node,  enabling comprehensive inference. 

\begin{table}[!h]
\centering
\caption{Comparison of Bayesian Network Structure Learning Methods learned using data-driven and LLM-enhanced learning methods.}
\vspace{-.4cm}
\label{tab:llm_improvement}
\footnotesize
\setlength{\tabcolsep}{3pt}
\begin{tabular}{lcccccc}
\toprule
\textbf{Method} &  \textbf{LLM} & \textbf{Correct} & \textbf{Miss} & \textbf{Rev.} & \textbf{Dir.} & \textbf{Recall}  \\
 &\textbf{Model} & $\uparrow$ & $\downarrow$ & $\downarrow$ &  $\uparrow$ & $\uparrow$ \\
\midrule
\multicolumn{7}{l}{\textit{LLM-Enhanced Methods}} \\
\midrule
{Hill climb + \gls{bic} + LLM } & Claude4.5 Sonnet & {10} & 2 &  2 & { 0.83} & {0.71}  \\
{Hill climb + \gls{bic} + LLM } & gpt-oss-120b & {10} & 2 &  2 & {0.82} & {0.68}  \\
Hill climb + K2 + LLM &Claude4.5 Sonnet & 10 & 2 &  2 & 0.85 & 0.71  \\
Hill climb + K2 + LLM &gpt-oss-120b & 9 & 3 &  2 & 0.78 & 0.61  \\
Hill climb + BDeu + LLM &Claude4.5 Sonnet & 9 & 2 &  3 & 0.75 & 0.64  \\
Hill climb + BDeu + LLM &gpt-oss-120b & 8 & 2 &  4 & 0.69 & 0.58  \\
\midrule
\multicolumn{7}{l}{\textit{Data-Only Baselines}} \\
\midrule
Hill climb + \gls{bic} (Data-only) &-& 5 & 4 &  5 & 0.50 & 0.36  \\
Hill climb + K2 (Data-only) &-& 3 & 4 &  7 & 0.30 & 0.21  \\
Hill climb + BDeu (Data-only) &-& 3 & 4 &  7 & 0.30 & 0.21  \\
PC algorithm &-& 8 & 0 &  6 & 0.57 & 0.57  \\
\bottomrule
\multicolumn{7}{l}{\footnotesize Ground truth: 14 edges. Average of 10 runs per configuration.} 
\end{tabular}
\end{table}
\begin{figure*}[t]
\centering

\begin{subfigure}[t]{0.48\textwidth}
    \centering
    \setlength\fwidth{0.55\linewidth}
    \setlength\fheight{0.35\linewidth}
\begin{tikzpicture}

\definecolor{darkgray176}{RGB}{176,176,176}
\definecolor{green}{RGB}{0,128,0}
\definecolor{lightgray204}{RGB}{204,204,204}
\pgfplotsset{every tick label/.append style={font=\scriptsize}}
\begin{axis}[
width=\fwidth,
height=\fheight,
legend cell align={left},
legend columns=2,
legend style={
  fill opacity=0.8,
  draw opacity=1,
  text opacity=1,
  at={(0.5,1.02)},
  anchor=south,
  draw=gray,
  font=\scriptsize
},
tick align=inside,
tick pos=left,
x grid style={darkgray176},
xmin=-0.81, xmax=7.11,
xtick style={color=black},
xtick={0.45, 3.15, 5.85},
xticklabels={$P_0$, $\Gamma$, {$P_0, \Gamma$}},
y grid style={darkgray176},
ylabel={UL Throughput [Mbps]},
ymajorgrids,
ymin=0, ymax=72.2540922451871,
ytick style={color=black},
font=\scriptsize
]
\draw[draw=black,fill=blue,opacity=0.8,semithick] (axis cs:-0.45,0) rectangle (axis cs:0.45,32.6089141856392);
\draw[draw=black,fill=green,opacity=0.8,semithick] (axis cs:0.45,0) rectangle (axis cs:1.35,42.6779201934704);
\draw[draw=black,fill=blue,opacity=0.8,semithick] (axis cs:2.25,0) rectangle (axis cs:3.15,32.6089141856392);
\draw[draw=black,fill=green,opacity=0.8,semithick] (axis cs:3.15,0) rectangle (axis cs:4.05,53.299194865811);
\draw[draw=black,fill=blue,opacity=0.8,semithick] (axis cs:4.95,0) rectangle (axis cs:5.85,32.6089141856392);
\draw[draw=black,fill=green,opacity=0.8,semithick] (axis cs:5.85,0) rectangle (axis cs:6.75,53.299194865811);
\path [draw=black, draw opacity=0.7, thick]
(axis cs:0,20.0875565209805)
--(axis cs:0,45.1302718502979);

\path [draw=black, draw opacity=0.7, thick]
(axis cs:0.9,33.616634411225)
--(axis cs:0.9,51.7392059757157);

\path [draw=black, draw opacity=0.7, thick]
(axis cs:2.7,20.0875565209805)
--(axis cs:2.7,45.1302718502979);

\path [draw=black, draw opacity=0.7, thick]
(axis cs:3.6,47.308778069539)
--(axis cs:3.6,59.289611662083);

\path [draw=black, draw opacity=0.7, thick]
(axis cs:5.4,20.0875565209805)
--(axis cs:5.4,45.1302718502979);

\path [draw=black, draw opacity=0.7, thick]
(axis cs:6.3,47.308778069539)
--(axis cs:6.3,59.289611662083);

\addplot [thick, black, opacity=0.7, mark=-, mark size=5, mark options={solid}, only marks, forget plot]
table {%
0 20.0875565209805
0.9 33.616634411225
2.7 20.0875565209805
3.6 47.308778069539
5.4 20.0875565209805
6.3 47.308778069539
};
\addplot [thick, black, opacity=0.7, mark=-, mark size=5, mark options={solid}, only marks, forget plot]
table {%
0 45.1302718502979
0.9 51.7392059757157
2.7 45.1302718502979
3.6 59.289611662083
5.4 45.1302718502979
6.3 59.289611662083
};
\draw (axis cs:0,46.1962557476142) node[
  scale=0.55,
  anchor=south,
  text=black,
  rotate=0.0
]{\bfseries 32.61};
\draw (axis cs:0.9,52.8051898730319) node[
  scale=0.55,
  anchor=south,
  text=black,
  rotate=0.0
]{\bfseries 42.68};
\draw (axis cs:2.7,46.1962557476142) node[
  scale=0.55,
  anchor=south,
  text=black,
  rotate=0.0
]{\bfseries 32.61};
\draw (axis cs:3.6,60.3555955593992) node[
  scale=0.55,
  anchor=south,
  text=black,
  rotate=0.0
]{\bfseries 53.30};
\draw (axis cs:5.4,46.1962557476142) node[
  scale=0.55,
  anchor=south,
  text=black,
  rotate=0.0
]{\bfseries 32.61};
\draw (axis cs:6.3,60.3555955593992) node[
  scale=0.55,
  anchor=south,
  text=black,
  rotate=0.0
]{\bfseries 53.30};
\addlegendimage{area legend, fill=blue, opacity=0.8, draw=black}
\addlegendentry{Default}
\addlegendimage{area legend, fill=green, opacity=0.8, draw=black}
\addlegendentry{Recomm.}
\end{axis}

\end{tikzpicture}%
\hfill
    \setlength\fwidth{0.55\linewidth}
    \setlength\fheight{0.35\linewidth}
\begin{tikzpicture}

\definecolor{darkgray176}{RGB}{176,176,176}
\definecolor{green}{RGB}{0,128,0}
\definecolor{lightgray204}{RGB}{204,204,204}
\pgfplotsset{every tick label/.append style={font=\scriptsize}}

\begin{axis}[
width=\fwidth,
height=\fheight,
legend cell align={left},
legend columns=2,
legend style={
  fill opacity=0.8,
  draw opacity=1,
  text opacity=1,
  at={(0.5,1.02)},
  anchor=south,
  draw=gray,
  font=\scriptsize
},
tick align=inside,
tick pos=left,
x grid style={darkgray176},
xmin=-0.81, xmax=7.11,
xtick style={color=black},
xtick={0.45, 3.15, 5.85},  
xticklabels={$P_0$, $\Gamma$, {$P0, \Gamma$}},
y grid style={darkgray176},
ylabel={UL BLER},
ymajorgrids,
ymin=0, ymax=0.78,
ytick style={color=black},
font=\scriptsize
]
\draw[draw=black,fill=blue,opacity=0.8,semithick] (axis cs:-0.45,0) rectangle (axis cs:0.45,0.293300332749562);
\draw[draw=black,fill=green,opacity=0.8,semithick] (axis cs:0.45,0) rectangle (axis cs:1.35,0.0869531318016929);
\draw[draw=black,fill=blue,opacity=0.8,semithick] (axis cs:2.25,0) rectangle (axis cs:3.15,0.293300332749562);
\draw[draw=black,fill=green,opacity=0.8,semithick] (axis cs:3.15,0) rectangle (axis cs:4.05,0.0961126254375729);
\draw[draw=black,fill=blue,opacity=0.8,semithick] (axis cs:4.95,0) rectangle (axis cs:5.85,0.293300332749562);
\draw[draw=black,fill=green,opacity=0.8,semithick] (axis cs:5.85,0) rectangle (axis cs:6.75,0.0961126254375729);
\path [draw=black, draw opacity=0.7, thick]
(axis cs:0,0)
--(axis cs:0,0.693489413750901);

\path [draw=black, draw opacity=0.7, thick]
(axis cs:0.9,0.0361878590823234)
--(axis cs:0.9,0.137718404521062);

\path [draw=black, draw opacity=0.7, thick]
(axis cs:2.7,0)
--(axis cs:2.7,0.693489413750901);

\path [draw=black, draw opacity=0.7, thick]
(axis cs:3.6,0.0537480062256308)
--(axis cs:3.6,0.138477244649515);

\path [draw=black, draw opacity=0.7, thick]
(axis cs:5.4,0)
--(axis cs:5.4,0.693489413750901);

\path [draw=black, draw opacity=0.7, thick]
(axis cs:6.3,0.0537480062256308)
--(axis cs:6.3,0.138477244649515);

\addplot [thick, black, opacity=0.7, mark=-, mark size=5, mark options={solid}, only marks, forget plot]
table {%
0 0
0.9 0.0361878590823234
2.7 0
3.6 0.0537480062256308
5.4 0
6.3 0.0537480062256308
};
\addplot [thick, black, opacity=0.7, mark=-, mark size=5, mark options={solid}, only marks, forget plot]
table {%
0 0.693489413750901
0.9 0.137718404521062
2.7 0.693489413750901
3.6 0.138477244649515
5.4 0.693489413750901
6.3 0.138477244649515
};
\draw (axis cs:0,0.699355420405892) node[
  scale=0.55,
  anchor=south,
  text=black,
  rotate=0.0
]{\bfseries 0.29};
\draw (axis cs:0.9,0.143584411176054) node[
  scale=0.55,
  anchor=south,
  text=black,
  rotate=0.0
]{\bfseries 0.09};
\draw (axis cs:2.7,0.699355420405892) node[
  scale=0.55,
  anchor=south,
  text=black,
  rotate=0.0
]{\bfseries 0.29};
\draw (axis cs:3.6,0.144343251304506) node[
  scale=0.55,
  anchor=south,
  text=black,
  rotate=0.0
]{\bfseries 0.10};
\draw (axis cs:5.4,0.699355420405892) node[
  scale=0.55,
  anchor=south,
  text=black,
  rotate=0.0
]{\bfseries 0.29};
\draw (axis cs:6.3,0.144343251304506) node[
  scale=0.55,
  anchor=south,
  text=black,
  rotate=0.0
]{\bfseries 0.10};
\addlegendimage{area legend, fill=blue, opacity=0.8, draw=black}
\addlegendentry{Default}
\addlegendimage{area legend, fill=green, opacity=0.8, draw=black}
\addlegendentry{Recomm.}
\end{axis}

\end{tikzpicture}
    \vspace{-.2cm}
    \caption{Throughput and BLER results for \texttt{Sc1}.}
    \label{fig:multi_optimization_sc1}
\end{subfigure}
\hfill
\begin{subfigure}[t]{0.48\textwidth}
    \centering
    \setlength\fwidth{0.55\linewidth}
    \setlength\fheight{0.35\linewidth}
\begin{tikzpicture}

\definecolor{darkgray176}{RGB}{176,176,176}
\definecolor{green}{RGB}{0,128,0}
\definecolor{lightgray204}{RGB}{204,204,204}
\pgfplotsset{every tick label/.append style={font=\scriptsize}}
\begin{axis}[
width=\fwidth,
height=\fheight,
legend cell align={left},
legend columns=2,
legend style={
  fill opacity=0.8,
  draw opacity=1,
  text opacity=1,
  at={(0.5,1.02)},
  anchor=south,
  draw=gray,
  font=\scriptsize
},
tick align=inside,
tick pos=left,
x grid style={darkgray176},
xmin=-0.81, xmax=7.11,
xtick style={color=black},
xtick={0.45, 3.15, 5.85},
xticklabels={$P_0$, $\Gamma$, {$P0,\Gamma$}},
y grid style={darkgray176},
ylabel={UL Throughput [Mbps]},
ymajorgrids,
ymin=0, ymax=72.2540922451871,
ytick style={color=black},
font=\scriptsize
]
\draw[draw=black,fill=blue,opacity=0.8,semithick] (axis cs:-0.45,0) rectangle (axis cs:0.45,48.5646033057851);
\draw[draw=black,fill=green,opacity=0.8,semithick] (axis cs:0.45,0) rectangle (axis cs:1.35,48.5646033057851);
\draw[draw=black,fill=blue,opacity=0.8,semithick] (axis cs:2.25,0) rectangle (axis cs:3.15,48.5646033057851);
\draw[draw=black,fill=green,opacity=0.8,semithick] (axis cs:3.15,0) rectangle (axis cs:4.05,51.8529567642957);
\draw[draw=black,fill=blue,opacity=0.8,semithick] (axis cs:4.95,0) rectangle (axis cs:5.85,48.5646033057851);
\draw[draw=black,fill=green,opacity=0.8,semithick] (axis cs:5.85,0) rectangle (axis cs:6.75,53.299194865811);
\path [draw=black, draw opacity=0.7, thick]
(axis cs:0,37.2546564821112)
--(axis cs:0,59.874550129459);

\path [draw=black, draw opacity=0.7, thick]
(axis cs:0.9,37.2546564821112)
--(axis cs:0.9,59.874550129459);

\path [draw=black, draw opacity=0.7, thick]
(axis cs:2.7,37.2546564821112)
--(axis cs:2.7,59.874550129459);

\path [draw=black, draw opacity=0.7, thick]
(axis cs:3.6,44.3956236282263)
--(axis cs:3.6,59.310289900365);

\path [draw=black, draw opacity=0.7, thick]
(axis cs:5.4,37.2546564821112)
--(axis cs:5.4,59.874550129459);

\path [draw=black, draw opacity=0.7, thick]
(axis cs:6.3,47.308778069539)
--(axis cs:6.3,59.289611662083);

\addplot [thick, black, opacity=0.7, mark=-, mark size=5, mark options={solid}, only marks, forget plot]
table {%
0 37.2546564821112
0.9 37.2546564821112
2.7 37.2546564821112
3.6 44.3956236282263
5.4 37.2546564821112
6.3 47.308778069539
};
\addplot [thick, black, opacity=0.7, mark=-, mark size=5, mark options={solid}, only marks, forget plot]
table {%
0 59.874550129459
0.9 59.874550129459
2.7 59.874550129459
3.6 59.310289900365
5.4 59.874550129459
6.3 59.289611662083
};
\draw (axis cs:0,60.9405340267752) node[
  scale=0.55,
  anchor=south,
  text=black,
  rotate=0.0
]{\bfseries 48.56};
\draw (axis cs:0.9,60.9405340267752) node[
  scale=0.55,
  anchor=south,
  text=black,
  rotate=0.0
]{\bfseries 48.56};
\draw (axis cs:2.7,60.9405340267752) node[
  scale=0.55,
  anchor=south,
  text=black,
  rotate=0.0
]{\bfseries 48.56};
\draw (axis cs:3.6,60.3762737976813) node[
  scale=0.55,
  anchor=south,
  text=black,
  rotate=0.0
]{\bfseries 51.85};
\draw (axis cs:5.4,60.9405340267752) node[
  scale=0.55,
  anchor=south,
  text=black,
  rotate=0.0
]{\bfseries 48.56};
\draw (axis cs:6.3,60.3555955593992) node[
  scale=0.55,
  anchor=south,
  text=black,
  rotate=0.0
]{\bfseries 53.30};
\addlegendimage{area legend, fill=blue, opacity=0.8, draw=black}
\addlegendentry{Default}
\addlegendimage{area legend, fill=green, opacity=0.8, draw=black}
\addlegendentry{Recomm.}
\end{axis}

\end{tikzpicture}%
    \hfill
   \vspace{-.2cm}
    \setlength\fwidth{0.55\linewidth}
    \setlength\fheight{0.35\linewidth}
\begin{tikzpicture}

\definecolor{darkgray176}{RGB}{176,176,176}
\definecolor{green}{RGB}{0,128,0}
\definecolor{lightgray204}{RGB}{204,204,204}

\begin{axis}[
width=\fwidth,
height=\fheight,
legend cell align={left},
legend columns=2,
legend style={
  fill opacity=0.8,
  draw opacity=1,
  text opacity=1,
  at={(0.5,1.02)},
  anchor=south,
  draw=gray,
  font=\scriptsize
},
tick align=inside,
tick pos=left,
x grid style={darkgray176},
xmin=-0.81, xmax=7.11,
xtick style={color=black},
xtick={0.45, 3.15, 5.85},  
xticklabels={$P_0$, $\Gamma$, {$P0, \Gamma$}},
y grid style={darkgray176},
ylabel={UL BLER},
ymajorgrids,
ymin=0, ymax=0.78,
ytick style={color=black},
font=\scriptsize
]
\draw[draw=black,fill=blue,opacity=0.8,semithick] (axis cs:-0.45,0) rectangle (axis cs:0.45,0.0852214876033058);
\draw[draw=black,fill=green,opacity=0.8,semithick] (axis cs:0.45,0) rectangle (axis cs:1.35,0.0852214876033058);
\draw[draw=black,fill=blue,opacity=0.8,semithick] (axis cs:2.25,0) rectangle (axis cs:3.15,0.0852214876033058);
\draw[draw=black,fill=green,opacity=0.8,semithick] (axis cs:3.15,0) rectangle (axis cs:4.05,0.0946872384937238);
\draw[draw=black,fill=blue,opacity=0.8,semithick] (axis cs:4.95,0) rectangle (axis cs:5.85,0.0852214876033058);
\draw[draw=black,fill=green,opacity=0.8,semithick] (axis cs:5.85,0) rectangle (axis cs:6.75,0.0961126254375729);
\path [draw=black, draw opacity=0.7, thick]
(axis cs:0,0)
--(axis cs:0,0.284894742743542);

\path [draw=black, draw opacity=0.7, thick]
(axis cs:0.9,0)
--(axis cs:0.9,0.284894742743542);

\path [draw=black, draw opacity=0.7, thick]
(axis cs:2.7,0)
--(axis cs:2.7,0.284894742743542);

\path [draw=black, draw opacity=0.7, thick]
(axis cs:3.6,0.0502594152726576)
--(axis cs:3.6,0.13911506171479);

\path [draw=black, draw opacity=0.7, thick]
(axis cs:5.4,0)
--(axis cs:5.4,0.284894742743542);

\path [draw=black, draw opacity=0.7, thick]
(axis cs:6.3,0.0537480062256308)
--(axis cs:6.3,0.138477244649515);

\addplot [thick, black, opacity=0.7, mark=-, mark size=5, mark options={solid}, only marks, forget plot]
table {%
0 0
0.9 0
2.7 0
3.6 0.0502594152726576
5.4 0
6.3 0.0537480062256308
};
\addplot [thick, black, opacity=0.7, mark=-, mark size=5, mark options={solid}, only marks, forget plot]
table {%
0 0.284894742743542
0.9 0.284894742743542
2.7 0.284894742743542
3.6 0.13911506171479
5.4 0.284894742743542
6.3 0.138477244649515
};
\draw (axis cs:0,0.286816995252294) node[
  scale=0.55,
  anchor=south,
  text=black,
  rotate=0.0
]{\bfseries 0.09};
\draw (axis cs:0.9,0.286816995252294) node[
  scale=0.55,
  anchor=south,
  text=black,
  rotate=0.0
]{\bfseries 0.09};
\draw (axis cs:2.7,0.286816995252294) node[
  scale=0.55,
  anchor=south,
  text=black,
  rotate=0.0
]{\bfseries 0.09};
\draw (axis cs:3.6,0.141037314223542) node[
  scale=0.55,
  anchor=south,
  text=black,
  rotate=0.0
]{\bfseries 0.09};
\draw (axis cs:5.4,0.286816995252294) node[
  scale=0.55,
  anchor=south,
  text=black,
  rotate=0.0
]{\bfseries 0.09};
\draw (axis cs:6.3,0.140399497158267) node[
  scale=0.55,
  anchor=south,
  text=black,
  rotate=0.0
]{\bfseries 0.10};
\addlegendimage{area legend, fill=blue, opacity=0.8, draw=black}
\addlegendentry{Default}
\addlegendimage{area legend, fill=green, opacity=0.8, draw=black}
\addlegendentry{Recomm.}
\end{axis}

\end{tikzpicture}

    \caption{Throughput and BLER results for \texttt{Sc2}.}
    \label{fig:multi_optimization_sc2}
\end{subfigure}
\vspace{-.1cm}
\caption{Multi-parameter optimization results based on tuning of parameters $P_0$, $\Gamma$, or $P_0$ and $\Gamma$ jointly.}
\label{fig:multi_optimization}
\end{figure*}
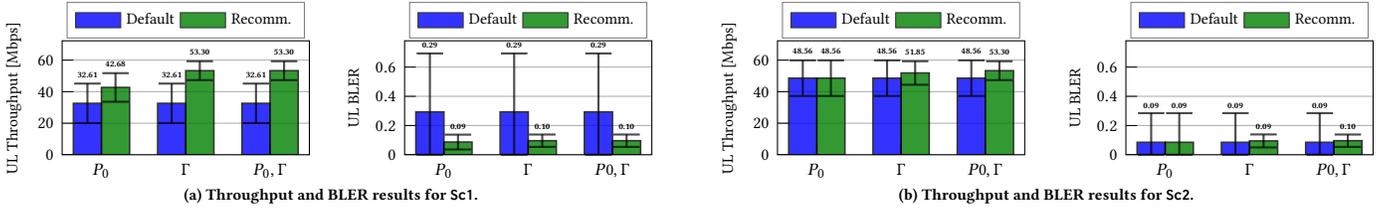
\begin{figure}[!t]
    \centering
        \setlength\fwidth{0.55\columnwidth} 
        \setlength\fheight{0.32\columnwidth} 
            \input{Figures/cdf}
                \vspace{-.4cm}
    \caption{\gls{cdf} of \glspl{kpi} for default and BN recommended parameter configurations.}
    \label{fig:cdf}
     \vspace{-.27cm}
\end{figure}

We evaluate the performance of the structures using metrics related to correctness and errors in the structure definition. Edges in the learned \gls{dag} can be classified into the following categories. \Circled{1} Correct edges: edges that are present in the ground truth, \Circled{2} Missed edges: edges that are present in the ground truth but not in the learned \gls{dag}, \Circled{3} Reversed edges: edges that are present in the ground truth and in the learned \gls{dag} but with incorrect direction, and \Circled{4} Extra edges: edges that are in the \gls{dag} but not in the ground truth. We compute the following metrics for the evaluation:
\begin{itemize}
    \item Directional accuracy is the fraction of edges with correct direction among ground truth edges that were found in the learned \gls{dag}, calculated as $\frac{Correct}{Correct + Reversed}$.
    \item Recall is the fraction of correct edges among all of the true edges ($\mathcal{T}$) calculated as $\frac{Correct}{\mathcal{T}}$.
\end{itemize}

We evaluate the learned \gls{dag} against a ground truth \gls{dag} subset with 14 edges representing high-confidence causal relationships, manually derived from 5G domain knowledge. We focus on this subset because, while 5G systems can contain broader causal networks, the inclusion of uncertain or context-dependent relationships would introduce ambiguity in the evaluation. We prioritize ground truth quality with unambiguous and protocol-validated edges, ensuring that high values for recall and directional accuracy indicate the discovery of established causal knowledge.  
We present the numerical results on the structural quality of the Bayesian Networks learned by data-driven and \gls{llm}-enhanced learning methods in Tab.~\ref{tab:llm_improvement}. We also compare multiple scoring functions, using the Hill Climb search algorithm with \gls{bic}, K2, and BDeu scores, and the PC algorithm~\cite{spirtes2000causation}. 


The table shows that LLM-enhanced models performs better than data-driven models. 
By incorporating domain knowledge, the best-performing method, Hill-climbing  enhanced by Claude 4.5 Sonnet using \gls{bic} and K2 scoring, recovers 71\% of the ground-truth edges compared to 36\% of the data-driven approach showcasing significant improvement. The results also shows that the LLM-integration aids in reducing reverse edges, resulting in better directional accuracy. We also note that the performance improvement is consistent across models using the Hill-climbing algorithm with different scoring functions.
These results shows that the DAG generated from the data with statistical method alone may not include causality and \gls{llm} aids in creating \gls{dag} structure with improved accuracy. This unlocks new data-driven applications for complex systems, once stymied by the lack of powerful structural learning solutions.

We also evaluate the performance of the LLM models Claude 4.5 Sonnet and gpt-oss-120b. 
The Bayesian Network structures learned by Claude 4.5 Sonnet slightly outperform gpt-oss-120b-derived structures across all scoring functions with the Hill climb search algorithm. 
However, while Claude achieves a slightly higher accuracy, gpt-oss-120b shows comparable results, with the advantage of local deployment. This enables network operators 
to perform LLM-enhanced optimization without using state-of-the-art closed models, like Claude, addressing data privacy 
concerns.
Based on these results, we use Claude 4.5 Sonnet as the \gls{llm} model and the Hill climb algorithm with \gls{bic} scoring as the base algorithm for structural learning of the Bayesian Network used in additional evaluations of this paper.

\subsection{Results on Configuration Optimization}

We evaluate the \framework framework by measuring the improvement of two critical uplink \glspl{kpi}: the throughput (\texttt{UL\_Mbps}) and the block error rate (\texttt{UL\_BLER}). We focus on the tuning of uplink power control parameters $P_0$ and $\Gamma$. Unless otherwise specified, each plot reports average throughput or BLER, and the whiskers represent the variance. 
\vspace{-0.5cm}
\paragraph*{Multi-Objective Optimization} First, we perform multi-objective optimization of single and multiple configuration parameters with throughput and BLER as target \glspl{kpi}. In the first scenario (\texttt{Sc1}), we use $P_0 = -90$~dBm and 
$\Gamma = 12$ dB, which is the default configuration provided by the \gls{oai} software stack for the Foxconn \gls{ru} we use in the private 5G network. In the second scenario (\texttt{Sc2}), we use $P_0 = -96$~dBm and $\Gamma = 28$ dB as an alternative default configuration. 



    

Figure~\ref{fig:multi_optimization} presents the  results for both the scenarios. From Fig.~\ref{fig:multi_optimization_sc1}, we observe that joint optimization in \texttt{Sc1} achieves a mean throughput of 53.30 Mbps, which is a 63.5\% improvement over the default configuration, and a mean block error rate of 9.6\%, a 19.7\% reduction from the baseline of 29.3\%. The optimization algorithm systematically evaluates all candidate configurations and selects the one with highest weighted expected performance as defined in Eq.~\eqref{eq:objective}. The configuration that the algorithm selected as optimal ($\mathbf{c}^*$) is $P_0 = -90$ dBm and $\Gamma = 20$ dB, with a score $S_{max}$ equal to 6.52. The second and third configurations with highest score reported are $P_0 = -84$ dBm with $\Gamma = 20$ dB and $P_0 = -106$ dBm with $\Gamma = 15$ dB. The performance of each of these configurations are: throughput 53.3 Mbps, BLER 9.6\%; throughput 51.26 Mbps, BLER 9.25\%; throughput 52.69 Mbps, BLER 9.46\%, respectively.
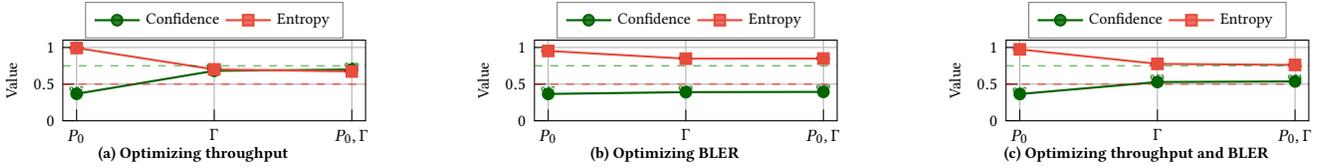
\begin{figure*}[tbp]
    \centering  
    \begin{subfigure}[t]{0.32\textwidth}  
        \centering
        \setlength\fwidth{0.95\linewidth}  
        \setlength\fheight{0.45\linewidth}  
\begin{tikzpicture}

\definecolor{darkgray176}{RGB}{176,176,176}
\definecolor{darkgreen}{RGB}{0,100,0}
\definecolor{darkred}{RGB}{139,0,0}
\definecolor{green}{RGB}{0,128,0}
\definecolor{lightgray204}{RGB}{204,204,204}

\definecolor{tomato2317660}{RGB}{231,76,60}

\begin{axis}[
width=\fwidth,
height=\fheight,
legend cell align={left},
legend columns=2,
legend style={
  fill opacity=0.8,
  draw opacity=1,
  text opacity=1,
  at={(0.5,1.02)},
  anchor=south,
  draw=gray,
  font=\scriptsize
},
tick align=inside,
tick pos=left,
x grid style={darkgray176},
xmajorgrids,
xmin=-0.1, xmax=2.1,
xtick style={color=black},
xtick={0,1,2},
xticklabels={$P_0$, $\Gamma$, {$P_0, \Gamma$}},
y grid style={darkgray176},
ylabel={Value},
ymajorgrids,
ymin=0, ymax=1.1,
ytick style={color=black},
font=\scriptsize
]
\addplot [thick, darkgreen, mark=*, mark size=2, mark options={solid}]
table {%
0 0.368507253446814
1 0.681294265567804
2 0.699081343462397
};
\addlegendentry{Confidence}
\addplot [thick, tomato2317660, mark=square*, mark size=2, mark options={solid}]
table {%
0 0.992885502632804
1 0.700923845849793
2 0.672939799023643
};
\addlegendentry{Entropy}
\addplot [semithick, green, opacity=0.5, dashed, forget plot]
table {%
-0.1 0.75
2.1 0.75
};
\addplot [semithick, red, opacity=0.5, dashed, forget plot]
table {%
-0.1 0.5
2.1 0.5
};
\draw (axis cs:0,0.398507253446814) node[
  scale=0.5,
  anchor=base,
  text=darkgreen,
  rotate=0.0
]{\bfseries 0.37};
\draw (axis cs:0,0.962885502632804) node[
  scale=0.5,
  anchor=base,
  text=darkred,
  rotate=0.0
]{\bfseries 0.99};
\draw (axis cs:1,0.711294265567804) node[
  scale=0.5,
  anchor=base,
  text=darkgreen,
  rotate=0.0
]{\bfseries 0.68};
\draw (axis cs:1,0.670923845849793) node[
  scale=0.5,
  anchor=base,
  text=darkred,
  rotate=0.0
]{\bfseries 0.70};
\draw (axis cs:2,0.729081343462397) node[
  scale=0.5,
  anchor=base,
  text=darkgreen,
  rotate=0.0
]{\bfseries 0.70};
\draw (axis cs:2,0.642939799023643) node[
  scale=0.5,
  anchor=base,
  text=darkred,
  rotate=0.0
]{\bfseries 0.67};
\end{axis}

\end{tikzpicture}
        \vspace{-.3cm}
        \caption{Optimizing throughput}
        \label{fig:conf_tp}
    \end{subfigure}
    \hfill
    \begin{subfigure}[t]{0.32\textwidth}  
        \centering
        \setlength\fwidth{0.95\linewidth}
        \setlength\fheight{0.45\linewidth}
\begin{tikzpicture}

\definecolor{darkgray176}{RGB}{176,176,176}
\definecolor{darkgreen}{RGB}{0,100,0}
\definecolor{darkred}{RGB}{139,0,0}

\definecolor{lightgray204}{RGB}{204,204,204}

\definecolor{tomato2317660}{RGB}{231,76,60}

\begin{axis}[
width=\fwidth,
height=\fheight,
legend cell align={left},
legend columns=2,
legend style={
  fill opacity=0.8,
  draw opacity=1,
  text opacity=1,
  at={(0.5,1.02)},
  anchor=south,
  draw=gray,
  font=\scriptsize
},
tick align=inside,
tick pos=left,
x grid style={darkgray176},
xmajorgrids,
xmin=-0.1, xmax=2.1,
xtick style={color=black},
xtick={0,1,2},
xticklabels={$P_0$, $\Gamma$, {$P_0, \Gamma$}},
y grid style={darkgray176},
ylabel={Value},
ymajorgrids,
ymin=0, ymax=1.1,
ytick style={color=black},
font=\scriptsize
]
\addplot [thick, darkgreen, mark=*, mark size=2, mark options={solid}]
table {%
0 0.366504311999695
1 0.391152733528956
2 0.39448323293247
};
\addlegendentry{Confidence}
\addplot [thick, tomato2317660, mark=square*, mark size=2, mark options={solid}]
table {%
0 0.952777231955562
1 0.847323465869493
2 0.848838094841458
};
\addlegendentry{Entropy}
\addplot [semithick, darkgreen, opacity=0.5, dashed, forget plot]
table {%
-0.1 0.75
2.1 0.75
};
\addplot [semithick, red, opacity=0.5, dashed, forget plot]
table {%
-0.1 0.5
2.1 0.5
};
\draw (axis cs:0,0.396504311999695) node[
  scale=0.5,
  anchor=base,
  text=darkgreen,
  rotate=0.0
]{\bfseries 0.37};
\draw (axis cs:0,0.922777231955562) node[
  scale=0.5,
  anchor=base,
  text=darkred,
  rotate=0.0
]{\bfseries 0.95};
\draw (axis cs:1,0.421152733528956) node[
  scale=0.5,
  anchor=base,
  text=darkgreen,
  rotate=0.0
]{\bfseries 0.39};
\draw (axis cs:1,0.817323465869493) node[
  scale=0.5,
  anchor=base,
  text=darkred,
  rotate=0.0
]{\bfseries 0.85};
\draw (axis cs:2,0.42448323293247) node[
  scale=0.5,
  anchor=base,
  text=darkgreen,
  rotate=0.0
]{\bfseries 0.39};
\draw (axis cs:2,0.818838094841458) node[
  scale=0.5,
  anchor=base,
  text=darkred,
  rotate=0.0
]{\bfseries 0.85};
\end{axis}

\end{tikzpicture}
        \vspace{-.3cm}
        \caption{Optimizing BLER}
        \label{fig:conf_bler}
    \end{subfigure}
    \hfill
    \begin{subfigure}[t]{0.32\textwidth}  
        \centering
        \setlength\fwidth{0.95\linewidth}
        \setlength\fheight{0.45\linewidth}
\begin{tikzpicture}

\definecolor{darkgray176}{RGB}{176,176,176}
\definecolor{darkgreen}{RGB}{0,100,0}
\definecolor{darkred}{RGB}{139,0,0}
\definecolor{green}{RGB}{0,128,0}
\definecolor{lightgray204}{RGB}{204,204,204}

\definecolor{tomato2317660}{RGB}{231,76,60}

\begin{axis}[
width=\fwidth,
height=\fheight,
legend cell align={left},
legend columns=2,
legend style={
  fill opacity=0.8,
  draw opacity=1,
  text opacity=1,
  at={(0.5,1.02)},
  anchor=south,
  draw=gray,
  font=\scriptsize
},tick align=inside,
tick pos=left,
x grid style={darkgray176},
xmajorgrids,
xmin=-0.1, xmax=2.1,
xtick style={color=black},
xtick={0,1,2},
xticklabels={$P_0$, $\Gamma$, {$P_0, \Gamma$}},
y grid style={darkgray176},
ylabel={Value},
ymajorgrids,
ymin=0, ymax=1.1,
ytick style={color=black},
font=\scriptsize
]
\addplot [thick, darkgreen, mark=*, mark size=2, mark options={solid}]
table {%
0 0.365403532369962
1 0.528835662726455
2 0.536792471681379
};
\addlegendentry{Confidence}
\addplot [thick, tomato2317660, mark=square*, mark size=2, mark options={solid}]
table {%
0 0.973852063771332
1 0.776405945339765
2 0.761969549407278
};
\addlegendentry{Entropy}
\addplot [semithick, green, opacity=0.5, dashed, forget plot]
table {%
-0.1 0.75
2.1 0.75
};
\addplot [semithick, red, opacity=0.5, dashed, forget plot]
table {%
-0.1 0.5
2.1 0.5
};
\draw (axis cs:0,0.395403532369962) node[
  scale=0.5,
  anchor=base,
  text=darkgreen,
  rotate=0.0
]{\bfseries 0.37};
\draw (axis cs:0,0.943852063771332) node[
  scale=0.5,
  anchor=base,
  text=darkred,
  rotate=0.0
]{\bfseries 0.97};
\draw (axis cs:1,0.558835662726455) node[
  scale=0.5,
  anchor=base,
  text=darkgreen,
  rotate=0.0
]{\bfseries 0.53};
\draw (axis cs:1,0.746405945339765) node[
  scale=0.5,
  anchor=base,
  text=darkred,
  rotate=0.0
]{\bfseries 0.78};
\draw (axis cs:2,0.566792471681379) node[
  scale=0.5,
  anchor=base,
  text=darkgreen,
  rotate=0.0
]{\bfseries 0.54};
\draw (axis cs:2,0.731969549407278) node[
  scale=0.5,
  anchor=base,
  text=darkred,
  rotate=0.0
]{\bfseries 0.76};
\end{axis}

\end{tikzpicture}
        \vspace{-.3cm}
        \caption{Optimizing throughput and BLER}
        \label{fig:conf_joint}
    \end{subfigure}
    \caption{Confidence and entropy results for different optimization objectives.}
    \label{fig:confidence}
\end{figure*}
\begin{figure*}[h]
\centering

\begin{minipage}[t]{0.65\textwidth}  
    \centering
    \setlength\fwidth{0.5\linewidth}  
    \setlength\fheight{0.25\linewidth}
\begin{tikzpicture}
\definecolor{darkgray176}{RGB}{176,176,176}
\definecolor{darkslateblue3064175}{RGB}{30,64,175}
\definecolor{lightgray204}{RGB}{204,204,204}
\definecolor{lightpink252165165}{RGB}{252,165,165}
\definecolor{lightskyblue147197253}{RGB}{147,197,253}
\definecolor{royalblue59130246}{RGB}{59,130,246}
\definecolor{tomato2396868}{RGB}{239,68,68}
\definecolor{darkred}{RGB}{139,0,0}
\begin{axis}[
width=\fwidth,
height=\fheight,
legend cell align={left},
legend columns=1,
legend columns=2,
legend style={
  fill opacity=0.8,
  draw opacity=1,
  text opacity=1,
  at={(0.5,1.02)},
  anchor=south,
  draw=gray,
  font=\scriptsize
},
tick align=inside,
tick pos=left,
x grid style={darkgray176},
xmin=-0.68, xmax=5.48,
xtick style={color=black},
xtick={0,0.9,1.8,3,3.9,4.8},
xticklabels={{Def}, {BN\\past}, {BN\\curr}, {Def}, {BN\\past}, {BN\\curr}},
xticklabel style={align=center},
y grid style={darkgray176},
ylabel={UL Throughput [Mbps]},
ylabel shift=-5pt,
ymajorgrids,
ymin=0, ymax=65.9641546091015,
ytick style={color=black},
font=\scriptsize
]
\draw[draw=black,fill=royalblue59130246,opacity=0.8,semithick] (axis cs:-0.4,0) rectangle (axis cs:0.4,32.6089141856392);
\draw[draw=black,fill=darkslateblue3064175,opacity=0.8,semithick] (axis cs:0.5,0) rectangle (axis cs:1.3,53.299194865811);
\draw[draw=black,fill=lightskyblue147197253,opacity=0.8,semithick] (axis cs:1.4,0) rectangle (axis cs:2.2,53.299194865811);
\draw[draw=black,fill=tomato2396868,opacity=0.8,semithick] (axis cs:2.6,0) rectangle (axis cs:3.4,23.0500694444444);
\draw[draw=black,fill=darkred,opacity=0.8,semithick] (axis cs:3.5,0) rectangle (axis cs:4.3,4.14417808219178);
\draw[draw=black,fill=lightpink252165165,opacity=0.8,semithick] (axis cs:4.4,0) rectangle (axis cs:5.2,24.2159459459459);

\draw (axis cs:0,33.6748980829555) node[
  scale=0.55,
  anchor=south,
  text=black,
  rotate=0.0
]{\bfseries 32.61};
\draw (axis cs:0.9,54.3651787631272) node[
  scale=0.55,
  anchor=south,
  text=black,
  rotate=0.0
]{\bfseries 53.30};
\draw (axis cs:1.8,54.3651787631272) node[
  scale=0.55,
  anchor=south,
  text=black,
  rotate=0.0
]{\bfseries 53.30};
\draw (axis cs:3,24.1160533417607) node[
  scale=0.55,
  anchor=south,
  text=black,
  rotate=0.0
]{\bfseries 23.05};
\draw (axis cs:3.9,5.210161979508) node[
  scale=0.55,
  anchor=south,
  text=black,
  rotate=0.0
]{\bfseries 4.14};
\draw (axis cs:4.8,25.2819298432622) node[
  scale=0.55,
  anchor=south,
  text=black,
  rotate=0.0
]{\bfseries 24.22};
\addlegendimage{area legend, fill=royalblue59130246, opacity=0.8, draw=black}
\addlegendentry{Center users}
\addlegendimage{area legend, fill=tomato2396868, opacity=0.8, draw=black}
\addlegendentry{Edge users}
\end{axis}
\end{tikzpicture}
    \hspace{0cm}  
    \setlength\fwidth{0.5\linewidth}  
    \setlength\fheight{0.25\linewidth}
\begin{tikzpicture}
\definecolor{darkgray176}{RGB}{176,176,176}
\definecolor{darkslateblue3064175}{RGB}{30,64,175}
\definecolor{lightgray204}{RGB}{204,204,204}
\definecolor{lightpink252165165}{RGB}{252,165,165}
\definecolor{lightskyblue147197253}{RGB}{147,197,253}
\definecolor{royalblue59130246}{RGB}{59,130,246}
\definecolor{tomato2396868}{RGB}{239,68,68}
\definecolor{darkred}{RGB}{139,0,0}
\begin{axis}[
width=\fwidth,
height=\fheight,
legend cell align={left},
legend columns=2,
legend style={
  fill opacity=0.8,
  draw opacity=1,
  text opacity=1,
  at={(0.5,1.02)},
  anchor=south,
  draw=gray,
  font=\scriptsize
},
tick align=inside,
tick pos=left,
x grid style={darkgray176},
xmin=-0.68, xmax=5.48,
xtick style={color=black},
xtick={0,0.9,1.8,3,3.9,4.8},
xticklabels={{Def}, {BN\\past}, {BN\\cur}, {Def}, {BN\\past}, {BN\\cur}},
xticklabel style={align=center},
y grid style={darkgray176},
ylabel={UL BLER},
ylabel shift=-5pt,
ymajorgrids,
ymin=0, ymax=0.40796534938704,
ytick style={color=black},
font=\scriptsize
]
\draw[draw=black,fill=royalblue59130246,opacity=0.8,semithick] (axis cs:-0.4,0) rectangle (axis cs:0.4,0.293300332749562);
\draw[draw=black,fill=darkslateblue3064175,opacity=0.8,semithick] (axis cs:0.5,0) rectangle (axis cs:1.3,0.0961126254375729);
\draw[draw=black,fill=lightskyblue147197253,opacity=0.8,semithick] (axis cs:1.4,0) rectangle (axis cs:2.2,0.0961126254375729);
\draw[draw=black,fill=tomato2396868,opacity=0.8,semithick] (axis cs:2.6,0) rectangle (axis cs:3.4,0.0877920138888889);
\draw[draw=black,fill=darkred,opacity=0.8,semithick] (axis cs:3.5,0) rectangle (axis cs:4.3,0.0921345205479452);
\draw[draw=black,fill=lightpink252165165,opacity=0.8,semithick] (axis cs:4.4,0) rectangle (axis cs:5.2,0.0838827027027027);

\draw (axis cs:0,0.299166339404553) node[
  scale=0.55,
  anchor=south,
  text=black,
  rotate=0.0
]{\bfseries 0.29};
\draw (axis cs:0.9,0.101978632092564) node[
  scale=0.55,
  anchor=south,
  text=black,
  rotate=0.0
]{\bfseries 0.10};
\draw (axis cs:1.8,0.101978632092564) node[
  scale=0.55,
  anchor=south,
  text=black,
  rotate=0.0
]{\bfseries 0.10};
\draw (axis cs:3,0.0936580205438801) node[
  scale=0.55,
  anchor=south,
  text=black,
  rotate=0.0
]{\bfseries 0.09};
\draw (axis cs:3.9,0.0980005272029364) node[
  scale=0.55,
  anchor=south,
  text=black,
  rotate=0.0
]{\bfseries 0.09};
\draw (axis cs:4.8,0.089748709357694) node[
  scale=0.55,
  anchor=south,
  text=black,
  rotate=0.0
]{\bfseries 0.08};
\addlegendimage{area legend, fill=royalblue59130246, opacity=0.8, draw=black}
\addlegendentry{Center users}
\addlegendimage{area legend, fill=tomato2396868, opacity=0.8, draw=black}
\addlegendentry{Edge users}
\end{axis}
\end{tikzpicture}
    \vspace{-.4cm}
    \caption{Performance comparison for cell-center and cell-edge users: 
    (left) throughput, (right) BLER. Def is default, BN past a BN-recommended configuration for a previous scenario or iteration, BN curr the one for the current network conditions.}
    \label{fig:absolute_performance}
\end{minipage}
\hspace{0.05cm} 
\begin{minipage}[t]{0.33\textwidth}  

    \centering
    \setlength\fwidth{1\linewidth}
    \setlength\fheight{0.5\linewidth}
\begin{tikzpicture}

\definecolor{darkgray176}{RGB}{176,176,176}
\definecolor{lightgray204}{RGB}{204,204,204}
\definecolor{steelblue31119180}{RGB}{31,119,180}
\definecolor{sage}{RGB}{171, 189, 159}

\begin{axis}[
width=\fwidth,
height=\fheight,
legend cell align={left},
legend columns=2,
legend style={
  fill opacity=0.8,
  draw opacity=1,
  text opacity=1,
  at={(0.5,1.02)},
  anchor=south,
  draw=gray,
  font=\scriptsize
},
tick align=inside,
tick pos=left,
x grid style={darkgray176},
xmin=-0.6, xmax=5.2,
xtick style={color=black},
xtick={0,0.9,1.8,3,3.9,4.8},
xticklabels={BN,Rul,Gr,BN,Rul,Gr},
y grid style={darkgray176},
ylabel style={align=center},
ylabel={UL Throughput\\Improvement [\%]},
ylabel shift=-3pt,
ymajorgrids,
ymin=-100, ymax=85,
ytick style={color=black},
font=\scriptsize
]

\draw[draw=black,fill=steelblue31119180,opacity=0.8,semithick] (axis cs:-0.4,0) rectangle (axis cs:0.4,63.5);    
\draw[draw=black,fill=steelblue31119180,opacity=0.8,semithick] (axis cs:0.5,0) rectangle (axis cs:1.3,0);       
\draw[draw=black,fill=steelblue31119180,opacity=0.8,semithick] (axis cs:1.4,0) rectangle (axis cs:2.2,63.5);    

\draw[draw=black,fill=sage,opacity=0.8,semithick] (axis cs:2.6,0) rectangle (axis cs:3.4,5.1);     
\draw[draw=black,fill=sage,opacity=0.8,semithick] (axis cs:3.5,0) rectangle (axis cs:4.3,-83.2);   
\draw[draw=black,fill=sage,opacity=0.8,semithick] (axis cs:4.4,0) rectangle (axis cs:5.2,-15.5);   

\draw (axis cs:0,66) node[scale=0.55, anchor=south, text=black]{\bfseries +64\%};
\draw (axis cs:0.9,2) node[scale=0.55, anchor=south, text=black]{\bfseries 0\%};
\draw (axis cs:1.8,66) node[scale=0.55, anchor=south, text=black]{\bfseries +64\%};

\draw (axis cs:3,8) node[scale=0.55, anchor=south, text=black]{\bfseries +5\%};
\draw (axis cs:3.9,-80) node[scale=0.55, anchor=north, text=black]{\bfseries -83\%};
\draw (axis cs:4.8,-13) node[scale=0.55, anchor=north, text=black]{\bfseries -16\%};

\addlegendimage{area legend, fill=steelblue31119180, opacity=0.8, draw=black}
\addlegendentry{Center users}
\addlegendimage{area legend, fill=sage, opacity=0.8, draw=black}
\addlegendentry{Edge users}
\end{axis}

\end{tikzpicture}
    \vspace{-.4cm}
    \caption{Percentage improvement over baseline for different 
    methods.}
    \label{fig:improvement_comparison}
\end{minipage}
\label{fig:context_adaptive}
\end{figure*}
From Fig.~\ref{fig:multi_optimization_sc2}, we observe that in \texttt{Sc2}, optimizing $P_0$ alone yields no improvement over the default (48.56 Mbps). Joint optimization  yields 53.30 Mbps throughput which is 9.7\% improvement over default and a 2.8\% improvement over the best single-parameter optimization. The \gls{bler} observed is 10.0\%, which is slightly higher than that of single parameter optimization. This suggest an inherent trade-off in multi-objective optimization, where throughput maximization with BLER minimization creates competing constraints. Despite this trade-off, the model balances both objectives achieving high throughput while maintaining an acceptable BLER rate. 

In the single configuration parameter optimization, we observe that throughput improvement is not drastic when optimized with power alone in both scenarios \texttt{Sc1} and \texttt{Sc2}. This is because with $\Gamma$ fixed, the achievable performance region is constrained to a limited subspace where $P_0$ adjustments yield minimal impact. We also observe that target SNR have a huge impact in throughput improvement in both scenarios and the we note that the performance is same as of joint optimization in scenario 2. This is because we fix $P_0$ to its default value during the single parameter optimization, and the joint optimization reveals that this default value is already optimal, leaving little room for improvement.

While the previous analysis focuses on average \gls{kpi} improvements, Fig.~\ref{fig:cdf} provides a more comprehensive view through empirical \glspl{cdf}, revealing the complete probability distributions under the default parameter configurations and the Bayesian Network recommended configurations. The \glspl{cdf} show consistent improvement of the recommended configurations over the default ones, across the entire distribution, not just at the mean.

\paragraph*{Prediction Uncertainty Analysis} A key advantage of the Bayesian Network approach over alternative methods is the uncertainty quantification, which includes confidence scores, indicating the prediction certainty, and entropy, indicating the spread of the distribution. This provides operators with interpretable reliability metrics.
Figure~\ref{fig:confidence} presents the prediction confidence of \texttt{Sc1} for the following optimization objectives:
(1) optimizing throughput, (2) optimizing \gls{bler}, and (3) jointly optimizing the throughput and \gls{bler} \glspl{kpi}. Results reveal key insights. The first insight is on parameter importance hierarchy. For example, in Fig.~\ref{fig:conf_tp}, we observe that optimizing $\Gamma$ alone achieves high confidence (0.70) for throughput prediction, while optimizing $P_0$ alone yields poor confidence (0.37), indicating that $\Gamma$ selection is the dominant factor in throughput determination among the two parameters considered. 
The second insight is on \gls{kpi}-dependent predictability. Comparing Figs.~\ref{fig:conf_tp} and~\ref{fig:conf_bler}, we observe that throughput predictions achieve confidence up to $0.70$, while BLER predictions reach only $0.37-0.39$ despite discretization of values. This reflects the high variance of BLER within the configuration. The reduced overall confidence also denotes the fact that BLER is weakly correlated with power parameters along with the tradeoffs between objectives. We observe that joint optimization for both \glspl{kpi} (Fig.~\ref{fig:conf_joint}) shows a moderate confidence (0.54), which is the direct consequence of high throughput prediction confidence and lower BLER prediction confidence.

\paragraph*{Deployment Across Contexts and Comparison with Baselines} The causal Bayesian Network can adapt to different deployment contexts. 
To show this capability, we consider a \gls{gnb} deployed to serve near, \gls{los} users with good \gls{rf} conditions, and a \gls{gnb} deployed to serve scenarios that include mostly \gls{nlos} users, obstructed indoor locations, or high-mobility users (e.g., vehicular users on highways, pedestrians in dense urban environments with multipath).
We refer to the first scenario as deployment with cell-center users, and the second with cell-edge users.  Figure~\ref{fig:absolute_performance} presents this capability. As in the previous experiment, we use $P_0 = -90$~dBm and 
$\Gamma = 12$ dB as the default configuration. To demonstrate that the configurations optimized for one context do not necessarily generalize to others, we also plot the KPI values using the configuration ($P_0=-90$ dBm, $\Gamma=20$ dB) that was recommended by the Bayesian Network in the previous experiment (BN past).
           

The model adapts the configuration recommendation for the cell-center users to $P_0=-90$ dBm and $\Gamma=20$ dB; and  cell-edge users to $P_0=-84$ dBm and $\Gamma=12$ dB. 
From Fig.~\ref{fig:absolute_performance}, we observe that both \glspl{kpi} improve for both the cell-center and cell-edge contexts. In particular, we note that the cell-edge users show modest improvement, as the performance is constrained by poor \gls{rf} conditions. We also observe that for cell-center users, the BN past recommendation performs well, showing that it is still appropriate for this scenario. However, for cell-edge users, the BN past recommendation significantly degrades performance compared to both the default and the current BN recommended configurations. This shows that a single static configuration cannot serve all user conditions effectively, necessitating context-aware optimization.
The ability of the Bayesian Network to produce context-specific recommendations emerges from learning the causal relationship between the variables. 

Finally, we evaluate the framework by comparing the recommendation against rule-based and greedy configuration adaptation methods. 
The rule-based baseline implements traditional engineering heuristics such as adjusting $P_0, \Gamma$ through fixed threshold mappings. These rules represent manual tuning performed by network engineers. The greedy method identifies the single configuration that achieved the best average throughput historically. From Fig.~\ref{fig:improvement_comparison}, we observe that the Bayesian recommendation consistently achieve better performance compared to baselines---default configuration, rule based and greedy across different user contexts. We note that our approach shows improvement in both throughput and BLER \glspl{kpi}. The greedy heuristics is based on the historical average and assumes that the data distribution remains constant. Consequently, it overlooks variability, so if there are configuration combinations have similar means but very different reliability, it may fail to distinguish between them. In addition, it fails at capturing dynamic behaviors, which are typical in networked systems, as we discuss above.  
When multiple configurations achieve similar mean performance, the \framework framework selects configurations with lower variance, providing more reliable and predictable performance.


We note that even though the evaluation is done with two configuration parameters, the approach is scalable: individual Bayesian Network structures can be learned for each parameter, then combined into a unified causal model through graph union operations. This modular construction enables scaling to arbitrary numbers of configuration parameters while managing computational complexity.


\paragraph*{Scalability Analysis} 
Table~\ref{tab:scalability} reports runtime measurements of the three components of \framework as the number of variables $|\mathcal{V}|$ increases from 10 to 22 which includes parameters and \glspl{kpi}. The profiling uses Claude-4.5-Sonnet.
The \gls{llm} constraint extraction takes 
46.7-108.5~s per invocation, scaling approximately linearly with the number of variables.
This is a one-time cost that does not execute during the operational control loop. 
The structure learning and configuration recommendation stages complete in under 6~seconds at all cases, well within the control loop timescales of the \gls{nrtric}. The \gls{llm}-derived constraints reduce the graph density, keeping treewidth manageable.
\begin{table}[t]
\centering
\vspace{.2cm}
\caption{Computational profiling of the \framework pipeline. }
\label{tab:scalability}
\vspace{-.3cm}
\begin{tabular}{lcccc}
\toprule
\textbf{$|\mathcal{V}|$} & \textbf{$|\mathcal{E}|$} & 
\textbf{LLM/run} (s) & \textbf{Structure} (s) & 
\textbf{Inference} (s) \\
\midrule
10 & 27 & 46.7   & 0.70  & 0.06 \\
19 & 51 & 74.9   & 3.98  & 0.16  \\
22 & 65 & 108.5  & 5.50  & 0.34  \\
\bottomrule
\end{tabular}
\vspace{-.3cm}
\end{table}


\section{Conclusion}
\label{sec:conclusions}

In this paper, we addressed the challenge of configuration parameter optimization in 5G networks. We proposed an \gls{llm}-enhanced Bayesian Network approach that integrates domain knowledge from 5G specifications with statistical structure learning. Our results show that large language models can effectively provide domain expertise for causal discovery in complex systems and thus enable better structure learning where statistical methods alone are insufficient. The experimental validations confirmed that configurations generated by the framework yield measurable performance improvements. Future directions of \framework include validation across multiple deployment 
scenarios such as urban versus rural morphologies, and indoor versus outdoor propagation environments.
\bibliographystyle{ACM-Reference-Format}
\bibliography{ref}
\end{document}